\documentclass[11pt,a4paper]{article}

\usepackage[a4paper,margin=2.5cm]{geometry}
\usepackage{authblk}

\usepackage{graphicx}
\usepackage{amsmath,amssymb}
\usepackage{xcolor}
\usepackage{microtype}
\usepackage{placeins}
\usepackage[numbers,sort&compress]{natbib}
\usepackage[hidelinks]{hyperref}
\graphicspath{{./}}
\newcommand{\keywords}[1]{\par\noindent\textbf{} #1}

\usepackage{caption}
\DeclareCaptionFont{figcapfont}{\fontsize{8pt}{10.8pt}\selectfont}
\AtBeginDocument{\captionsetup[figure]{font=figcapfont,labelfont=bf}}

\begin{document}

%\title{The Information Contained in Time-Derivatives Spaces}
\title{Physically-Relevant Information Learning in High-Dimensional Time-Derivatives Spaces}
% Building and navigating HDTDS 
% The information contained in HDTDS 
% Extracting information from HDTDS 
% Information search into HDTDS 

\author[1]{Domiziano Doria}
\author[1]{Matteo Becchi}
\author[1]{Giovanni M. Pavan\thanks{Corresponding author: giovanni.pavan@polito.it}}
\affil[1]{Department of Applied Science and Technology, Politecnico di Torino, Torino 10129, Italy}
\date{\today}

\maketitle

\begin{abstract}
Understanding the physics of many-body complex dynamical systems may be a non-trivial task. High-dimensional analysis approaches are often deemed necessary to prevent losing important information. Typically, these use order parameters or descriptors capturing information related to, e.g., relative positions, symmetries, etc., of the units in the studied system. However, in many cases, gaining information related to the relative positions of the constitutive units (or their velocities) alone may be insufficient, and to reach a more complete physical knowledge, one should ideally learn and correlate with each other both structure and dynamics. Here we demonstrate how to achieve such a goal efficiently by building and navigating high-dimensional Time-Derivatives (TiDe) spaces. A TiDe space can be generated for virtually any type of system/phenomenon from the time-series data collected along its observation over time. {Each TiDe’s dimension corresponds to a growing-order time-derivative of the extracted data, thus containing information related to different physical phenomena/events, which can be easily extracted via unsupervised approaches.} {We demonstrate how, by definition, TiDes can be directly analyzed without a need for prior dimensionality reduction, providing results that are intrinsically intuitive to interpret.} We show the potential of the method by analyzing two prototypical example datasets extracted from molecular dynamics simulations or experimental tracking of different types of complex dynamical systems. Our results demonstrate how efficiently one can navigate and learn in information-rich TiDe spaces, which provide a robust general framework for data analysis and for studying complex dynamical systems from the data collected along their observation over time.
%Our results demonstrate how efficiently one can navigate and learn in the space of time derivatives, which provides a robust paradigm useful for the study of any type of complex system (phenomenon) from its time-series data, as well as for data-analysis in general.
\end{abstract}

%\keywords{Time derivatives $|$ High-dimensional space $|$ Complex systems $|$ Time-series $|$ Information extraction}
\newpage
%\section*{Significance Statement}
%The purpose of high-dimensional analysis is reaching a comprehensive understanding of a system by integrating different parameters (dimensions) ideally complementary and independent from each other. However, designing efficient high-dimensional analyses is most often non-trivial, especially when combining purely structural descriptors/parameters. Here we demonstrate how information-rich analyses can be attained by moving along the “time-derivatives dimensions”. For any type of data acquired over time, a high-dimensional time-derivatives (TiDe) space can be created which dimensions are the growing-order time-derivatives of the original time-series data. Navigation of such TiDes and their dimensions allows extracting physically-interpretable information on different types of phenomena in efficient way. This offers a precious platform for studying complex dynamical systems and for unveiling relevant information nested in time-series data.

\section*{Introduction}
Deciphering the behavior of complex systems through their observation over time can be a significant challenge. This is particularly true for many-body systems, whose behaviors and properties originate from the intricate networks of dynamical interactions between the many units composing them.~\cite{Crippa2025-vi,crippa2022molecular,hallstrom2026decoding} Reaching a profound characterization of the internal physics of such systems often requires tracking the individual constitutive units in the system and translating their reciprocal behaviors (deep-encoded into their raw trajectories) into meaningful and interpretable data by means of observables, parameters, or descriptors, chosen to study the system. 
Typically, one can choose between different types of observables (or descriptors): from the most abstract and general, to heuristic ones based on the experience and prior knowledge of the system under study.~\cite{musil2021physics,crippa2023detecting,martino2024data,caruso2023timesoap} 

Nonetheless, in general, any descriptor is, on its own, by definition incomplete to some extent, and can effectively capture some key features of the system under study while overlooking others.~\cite{donkor2023machine, martino2024data,Crippa_2023_MLST,caruso_pnas_nexus_classification} 

For this reason, high-dimensional analyses are typically deemed desirable, if not necessary, to reduce the risk of losing important features of the studied system and to enrich the analysis.
Such high-dimensionality can be obtained in various ways. One typical approach is to combine different types of descriptors in the attempt to capture different events in a multi-parametric analysis. However, most often it is non-obvious to what extent the various parameters complement each other or provide redundant information.~\cite{donkor2023machine, martino2024data,lionello2025relevant}
Another approach is to use descriptors that are inherently (by definition) high-dimensional~\cite{bartok2013representing, drautz2019atomic}. Among many~\cite{drautz2019atomic, nigam2020recursive}, one example descriptor widely used to study and classify atomic/molecular structures is the Smooth Overlap of Atomic Positions (SOAP),~\cite{bartok2013representing} which provides a high-dimensional embedding useful for learning the structural motifs surrounding every SOAP center (molecule or atom) revealing the populated/predominant ones in the studied system~\cite{rowe2020accurate, monserrat2020liquid, capelli2022ephemeral, capelli2021data, gabellini2022spotting, rapetti2023machine, sormani2025opportunities,cioni_metals_2023,gasparotto_2020}.
However, while such approaches benefit from a feature space that is intrinsically high-dimensional (each component being mathematically orthogonal to the other ones), most often the various dimensions contain just different projections of the very same physical phenomena ~\cite{lionello2025relevant}. 
On a practical level, these issues may provide the erroneous impression to tailor a complete high-dimensional analysis, while in reality the information captured by such a large number of dimensions (or parameters) might be very low (even mono) dimensional and refer to  the same individual physical phenomenon.~\cite{lionello2025relevant} In such common cases, it has been also recently demonstrated how adding more-and-more parameters (or descriptors/dimensions) that are ``redundant'' but little informative to each other may not only be useless, but also detrimental for information extraction due to recurrent negligible-information/relevant-noise additions and information-frustration issues.~\cite{altman2018curse, lionello2025relevant, becchi2025maximum} 
As a corollary, the methods that are then subsequently chosen to reduce the dimensionality in such analyses may considerably change the obtained results, often making it non-trivial to rationalize and verify them creating a sort of black-box analysis.~\cite{gasparotto_2020,capelli2021data,ChengB_2020_mapping,ceriotti2019_unsupervised,Deringer_2020_phosphorus,hassanali_Sormani_2025_unsupervised,caruso2023timesoap,lionello2025relevant,becchi2025maximum,Ulugol_Filion_comparing_unsuper_2025} 

Similar limitations to those described above may typically arise when analyzing a system by combining, many, but purely structural, features.~\cite{lionello2025relevant,donkor2023machine} On the other hand, considerable benefits have been reported when combining descriptors of local structure and dynamics together, as these may effectively capture events that are in some sense really "orthogonal" to each other.~\cite{Crippa_2023_MLST,caruso_pnas_nexus_classification}
This indicates how it might be more efficient to combine fewer descriptors that can effectively capture both the structural and dynamical features of the system under study rather than, e.g., combining a large number of order parameters, in a very similar way as, for reaching a comprehensive understanding of the physics of a dynamical system, one should ideally obtain information on both positions and velocities.

% REVISION 
Time-derivatives of reference data are known to contain useful information that can be used to reconstruct the system's dynamics: successive derivatives of, e.g., displacement have clear physical meanings, such as velocity, acceleration, jerk, snap, etc.\cite{eager2016beyond,packard1980geometry} Enriching instantaneous observations with temporal information is a concept also used by methods that learn dynamical models from data: as reference, sparse identification of nonlinear dynamics (SINDy) uses, e.g., measured variables together with their rates of change to identify equations describing how a system evolves.\cite{brunton2016sindy} Hankel alternative view of Koopman (HAVOK) analysis instead uses a sequence of current and past measurements of a signal, representing chaotic dynamics through a linear model with an additional forcing term associated with intermittent changes in behavior.\cite{brunton2017havok}  
%Al such approaches indicate how following observations over time can reveal information about the underlying dynamics. Their aim is to model system evolution, whereas here we use successive descriptor derivatives to distinguish local environments and transitions. 
In molecular simulations analysis, another example is given by TimeSOAP: a SOAP-based\cite{bartok2013representing} descriptor that tracks time-changes in the SOAP spectrum of each molecule over time\cite{caruso2023timesoap}. The time derivative to the Time-SOAP signal has been demonstrated to provide interesting insights, e.g., on the domains that are undergoing local structural rearrangements in dynamical many-body molecular systems of different nature.\cite{caruso2023timesoap,caruso_pnas_nexus_classification} All such previous works offer an interesting perspective, which is the starting point of this work.

Building on such fundamental concepts, here we describe an efficient method to maximize the extraction of physically-relevant information in the study of virtually any type of system from the data extracted during its observation over time. Our approach is based on the construction of high-dimensional Time-Derivatives (TiDe) spaces that may then be easily analyzed. Starting from the time-series data extracted from a system (or phenomenon) under study, a high-dimensional TiDe space can be readily constructed, the components of which are the growing-order time derivatives of the original time-series data. Each dimension in such TiDe spaces contains data that are the time-derivatives of those of the previous dimension, thus encoding information referring to different types of physical phenomena in the systems (e.g., displacements, velocities, accelerations, jerks, snaps, etc.). 

In particular, we show how this offers innate interpretability of the information extractable by each dimension of a TiDe space, and allows capturing both spatial/structural and dynamical/temporal information that are intrinsically complementary to each other in efficient way and without the need for invasive dimensionality reduction. 
The amount and complementarity of the information that is added into the analysis when integrating the successive time-derivative data contained in the various TiDe's dimensions can be systematically quantified in different ways.
We analyze different types of datasets extracted from different-scale prototypical example many-body dynamical systems, the trajectories of which are obtained from molecular dynamics simulations or experimental tracking. The results that we discuss herein demonstrate the potential, robustness, and transferability of this method, which we expect will provide a useful framework for the analysis of time-series data and for the study of any type of complex physical system from the data obtained during its observation over time.

\section*{Results}
\subsection*{Theoretical foundation}
The construction of the high-dimensional time-derivatives space (TiDe) begins from a univariate time-series data $D(t)$, measured or computed over a sequence of time frames $t$, and by systematically calculating the time-derivative of the data as:

\begin{equation}
    \dot{D}(t) \equiv \frac{D(t + \delta t) - D(t)}{\delta t}
    \label{eq:eq1}
\end{equation}

where $\delta t$ is the time step between consecutive frames.
The TiDe representation of maximum derivative order $n$ is thus composed of $n+1$ coordinates, corresponding to the original time-series and its successive time derivatives:

\begin{equation}
    \mathbf{TiDe} = (D(t), \dot{D}(t), \ddot{D}(t), \dots, D^{(n)}(t))
    \label{eq:eq2}
\end{equation}

Such raw signals typically contain both meaningful features and high-frequency fluctuations arising from thermal noise or numerical artifacts, which tend to increase at each successive derivation step. To prevent these fast oscillations from obscuring the meaningful information, before derivation over time, each time-series is first denoised using a low-pass Butterworth filter~\cite{butterworth1930theory} (other similar noise reduction approaches can be used to this end). 
This step removes spurious high-frequency components while preserving slower collective dynamics of physical relevance, thereby ensuring stability and robustness.
The scheme in the construction of the TiDe is thus: (i) compute $D^0\equiv D$, (ii) noise-filter the $D^0$ time-series, which becomes the first TiDe coordinate, (iii) compute the first time derivative $D^{(1)}\equiv\dot{D}$, which becomes the second TiDe coordinate, and (iv) iterate the derivation step up to the maximum derivative order $n$, obtaining $D^{(n)}$ as the $(n+1)^{th}$ TiDe dimension:

\begin{equation}
    D^{(n)}(t) \equiv \frac{D^{(n-1)}(t + \delta t) - D^{(n-1)}(t)}{\delta t}
    \label{eq:eq3}
\end{equation}

Further details on preprocessing, differentiation, and alignment of the derivative time-series are provided in the Materials and Methods, subsection ``Time- series filtering: preprocessing and low-pass filter.'' The resulting collection of time-series spans an $n+1$-dimensional space (the {\it embedding dimension}), which allows capturing and separating both key structural and dynamical features of the system and can then be directly exploited by clustering algorithms~\cite{liao2005clustering}. In the case of a many-body system composed of \textit{i} units that interact and move over time, for each of which a trajectory is available that can be converted into a univariate time-series using an arbitrary local descriptor (e.g., a local order parameter, etc.), this generates \textit{i} time-series: $D_i(t)$. In this case, each dimension of the TiDe space is composed of $i$ time-series, while for each unit $i$, the successive TiDe dimensions are the growing-order time derivatives of the original time-series data. Details of the trajectories, descriptors, and derivative calculations used below are provided in the Materials and Methods.

% REVISION 
The choice of descriptor determines the physical information that is practically captured in the analysis and that is thus contained in the TiDe's dimensions. A TiDe can be built starting from different types of time-series data. In the first examples shown here below, we use as an example a descriptor, such as, e.g., LENS,\cite{crippa2023detecting} which can efficiently track changes in local neighbors and can easily discriminate micro-domains with different local dynamicity, distinguishing for example ice from liquid water in ice/water coexistence simulations in unambiguous and unbiased way.\cite{crippa2023detecting} As it is demonstrated and discussed below, the TiDe approach described herein can be applied to virtually any type of data obtained with differentiable descriptors, including those capturing purely structural features or changes.\cite{martino2024data,caruso2023timesoap} The example reported at the end of this work, which focuses on a macroscopic active matter system, uses a different descriptor, well suited to capture collective dynamical events in complex dynamical ensembles: that is, the alignment of the velocities of neighbor particles, $\varphi$.

It is worth noting that in our method, the theoretical dimensionality of the TiDe space and the "real" dimensionality of the system the data refer to in the TiDe space are two different things. In the analysis, the maximum number of time-derivatives ($n$) that are iteratively calculated starting from the original data is pre-selected by the user. This sets the theoretical dimensionality of the TiDe space ($n+1$, from from $D^0$ to $D^{n}$). This of course does not imply that the system that will be studied will be really $n$-dimensional in this TiDe space (this would be the case if all $n$-dimensions contain complementary information in them).
In particular, from a practical point of view, extracting the information contained in each individual TiDe coordinate, and estimating the information add-on provided by adding the successive derivatives (TiDe dimensions) in the analysis, will allow assessing the effective dimensionality of the system in the TiDe space (see next sections). Thus, in this approach, what is important is just that $n$ is an overestimation of the real dimensionality of the system in the TiDe space: if additional/complementary information are gained only  only up to a certain derivative order $d$ (with $d<n$), then this simply means that the system under study is $d+1$-dimensional in the TiDe's space: i.e., only TiDe's dimensions from $D^0$ to $D^{d}$ contain useful information, while this can be quantified in our method in a straightforward way.

Recently, using an information-theory-based method based on the clustering-induced minimization of Shannon entropy called Maximum Information Extraction (MInE),\cite{becchi2025maximum} we demonstrated the tight relationship existing between discovering statistically relevant clusters in the data and the information that can be effectively extracted from the data. 
For a detailed explanation of the MInE method, we refer to the original MInE paper.\cite{becchi2025maximum} In brief, the Shannon entropy of an observable $x$, which takes values in a set $\chi$ with probability distribution $p(x)$, is defined as:

\begin{equation}
    H(x) = -\sum_{x \in \chi} p(x) \log_2 p(x)    
    \label{eq:eq4}
\end{equation}

Starting from time-series data, it is possible to calculate the (global) Shannon entropy of the data distribution, $H_0$.
When $n$ different micro-clusters can be distinguished in the data, with probability distributions $p_n(x)$, the Shannon entropy of the data after clustering can estimated as the weighted sum of the Shannon entropies of the micro-clusters\cite{becchi2025maximum}:

\begin{equation}
    H_{clust}(x) = \sum_{i=1}^{n} w_nH_n    
    \label{eq:eq5}
\end{equation}

where $w_1,..., w_{n}$ are the fraction of data belonging to the various clusters (i.e., their populations).
By concavity, the subdivision of the data into microclusters reduces entropy\cite{becchi2025maximum}. This means that when different micro-clusters are classified in the data, this produces a post-clustering Shannon entropy that is lower than that of the whole distribution (equal in the limit case where no distinct microclusters are present in the data). The information gained by unveiling the internal micro-clusters in the data (i.e., the data microstructure)\cite{baldo2026microinformation} can be then quantified in $[bit]$ as the difference between the Shannon entropy of the data pre- vs. post-clustering:

\begin{equation}
    \Delta I = H_{0}-H_{clust} \geq 0    
    \label{eq:eq6}
\end{equation}

As an efficient method to discover micro-clusters in time-series data (even hidden, but statistically-relevant ones), in the examples analyzed herein, we use Onion clustering~\cite{becchi2024layer}. 
Applied iteratively to datasets of increasing dimensionality, Onion clustering identifies in unsupervised way all dynamically-diverse micro-environments and transient states that can be effectively statistically discriminated even in high-dimensional representations, providing in output the number of statistically-different micro-domains that can be distinguished as a function of the time-resolution used in the time-series analysis.\cite{becchi2024layer,becchi2025maximum} 
However, we underline that any other type of method capable of providing a meaningful data clustering is useful to the quantification of the information contained in the data.\cite{baldo2026microinformation} 

Here, we use these approaches to quantify the information gained by analyzing all TiDe's dimensions and, consequently, the information add-on obtained by considering successive TiDe dimensions (and thus, time-derivatives) in the analysis, determining the effective dimensionality of the system under study in the TiDe's space.
TiDes effectively embed the intrinsic dynamical complexity of the system (and the information contained in it), originally sampled using e.g. a one-dimensional descriptor, into a high-dimensional space, which successive dimensions are one the time-derivative of the previous one. The approach is mathematically very similar to time-delay embedding in dynamical systems and chaos theory (Takens's embedding theorem),~\cite{noakes1991takens, takens2006detecting} albeit with some notable differences. 

First, by employing time derivatives instead of time-delayed signals, each dimension in the TiDe space has a clear and different physical meaning (referring to, e.g., changes, rates of change, accelerations, jerks, snaps, etc., in the data). 
{Each higher-order TiDe derivative dimension thus contains data that are the rate of change of the previous one: transitions between states detected in one time-series become themselves real metastable states in its derivative counterpart, which carry information that can be systematically quantified, classified, and physically interpreted.} 
{Through this construction, the space inherently encodes a hierarchy of temporal scales, from slow reorganizations to rapid fluctuations, enabling the detection of both smooth and abrupt dynamical transitions.}The resulting architecture offers a richer analytical framework for describing and distinguishing complex dynamical states, which becomes very instructive even without the need for invasive dimensionality reduction approaches, as is demonstrated below.

\subsection*{Example 1: Extracting information from ice-water coexistence MD trajectories}
\begin{figure}[ht!]
 \centering
 \includegraphics[width=\textwidth,keepaspectratio]{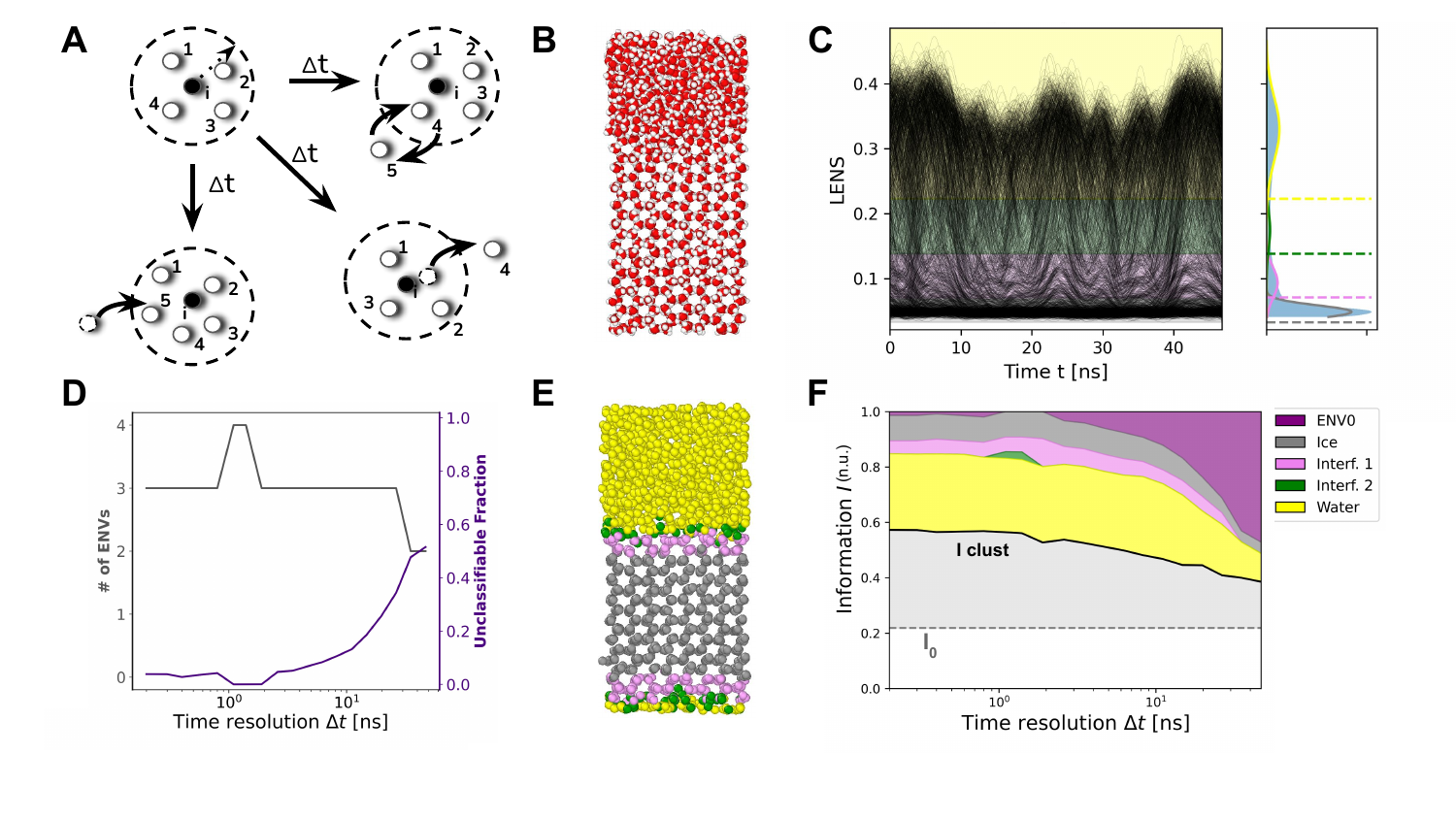}
 \caption{\textbf{Information extraction from LENS time-series data extracted from MD simulation trajectories of ice/water coexistence. } 
 (A) Schematic representation of the LENS descriptor,\cite{crippa2023detecting} which captures the local dynamics in the neighborhood of a central particle, tracking molecular neighbors' reshuffling within a specified cutoff radius (10~\AA\ in this case) over consecutive time frames (further details in the Methods section). 
 (B)Snapshot of the simulated MD system analyzed, consisting of 2048 TIP4P/ICE molecules coexisting in equilibrium between ice and liquid phases (simulation details provided in the Methods section). 
 (C) LENS time-series data for the 2048 water molecules along 50 ns of MD simulation. The background is colored based on the four main micro-clusters that are detected by analyzing the time-series with Onion clustering\cite{becchi2024layer} (using a time-resolution of $\Delta t=1.4$ ns: the time resolution granting maximum information extraction\cite{becchi2025maximum}). The various detected LENS environments (gaussian micro-clusters) are shown in the KDE (right) as colored solid gaussian curves, the inter-cluster thresholds are identified by the horizontal dotted lines (LENS signals have been processed using a Butterworth low-pass filter to remove high-frequency noise, see details in Methods section). 
 (D) Onion plot showing the number of statistically-relevant different micro-clusters (LENS environments: gray line, primary Y axis) that can be resolved by Onion clustering as a function of the time-resolution used to analyze the time-series data and the fraction of data points that cannot be resolved for insufficient time-resolution (purple line, secondary Y axis).\cite{becchi2024layer} Three microclusters referring to liquid water, solid ice, and the solid/liquid interface can be stably resolved down to a resolution of $\Delta t \lesssim 20$ ns. 
 (E) Snapshot from panel B with molecules color-coded according to the LENS environments of belonging detected at the selected resolution of $\Delta t=1.4$ ns,\cite{doria2025data,becchi2024layer,becchi2025maximum} where a maximum of four environments can be resolved - bulk ice (gray), ice interface (pink), liquid interface (green), and bulk liquid (yellow).
 (F) MInE plot showing the Information $I$ obtained by resolving the LENS micro-clusters with Onion clustering as a function of the time resolution $\Delta t$ used in the analysis (I in normalized units, n.u.).\cite{becchi2025maximum} The horizontal dashed line ($I_0$) represents the information content that can be calculated directly from the Shannon entropy of the data distributions. The solid black curve ($I_{clust}$) represents the total information obtained after clustering (i.e., after the discovery of the micro-clusters and the subdivision of the water molecules into them)\cite{becchi2025maximum} as a function of $\Delta t$: the gray shaded area corresponds to the information gain achieved via clustering of the data. In essence, the plot indicates how much of the data (in fraction) can be translated into information (at best, $I_{clust}\sim 58 \%$ in this case) and how much is entropy ($\sim 42 \%$ at the optimal resolution). The widths of the colored regions above the $I_{clust}$ black line reflect the relative weighted Shannon entropy of each environment. }
 \label{fig:fig1}
\end{figure}

As a first example, we analyze a prototypical Molecular Dynamics (MD) simulation of a system where solid ice and liquid water coexist in dynamical equilibrium in correspondence of the transition temperature. 
The simulation consists of $N=2048$ TIP4P/ICE\cite{abascal2005potential} water molecules -- half of which are initially in crystalline hexagonal ice phase, the other half in the liquid state (see Methods for simulation details) -- that we then analyze using the Local Environment and Neighbor Shuffling (LENS) descriptor (Fig.~\ref{fig:fig1}A,B).~\cite{crippa2023detecting}  
This is an example benchmark system in equilibrium with non-trivial internal dynamics.\cite{caruso2023timesoap,crippa2023detecting,Crippa_2023_MLST,becchi2024layer,becchi2025maximum} 
We analyze a 50~ns pre-equilibrated MD trajectory generated in the $NPT$ ensemble, during which water molecules are seen to dynamically exchange between the ice and liquid phase.~\cite{crippa2023detecting} 
We use the LENS as an efficient descriptor capable of capturing changes in the "list" of molecular individuals that surround each molecule in the system over time (at each timestep): it allows discriminating additions, losses, or permutations of neighbor molecules around every (water) molecule in the system, that is, around each LENS center (Fig.~\ref{fig:fig1}A). 
From the MD trajectories, we thus obtain 2048 individual LENS time-series (one for each water molecule in the system), using as the LENS centers the oxygen atom of every molecule in the system. We compute the LENS time-series by tracking permutations, additions, and removals of neighboring particles within a fixed cutoff radius $r_\text{cut} = 10$~\AA\ and by segmenting the total simulation trajectories in time-intervals of $\delta t = 100$~ps (these setup parameters are a very good fit to optimize information extraction in such a system, as discussed in detail in Refs.\cite{doria2025data,becchi2024layer,becchi2025maximum}, see also Methods). 

Each of the $2048$ LENS time-series contains one data point for each of the analyzed $T=500$ frames. 
As it can be expected in MD trajectories, the signals contain high-frequency fluctuations due to molecular vibrations. To mitigate noise and emphasize slow dynamics, we applied a zero-phase Butterworth low-pass filter~\cite{butterworth1930theory} to all time-series (filter parameters provided in Methods section). This initial noise-filtering is essential for downstream clustering analysis (as noise tends to increase at each successive derivation iteration). 
The noise-filtered LENS signals reveal two main populations (Fig.~\ref{fig:fig1}C, right: two gray and yellow histogram maxima): one narrower at lower LENS values (molecules vibrating in the crystalline ice lattice) and a second wider one at higher LENS values (dynamic particles in the liquid phase).

To detect all statistically-relevant dynamically-distinct micro-domains (environments) that populate the system, we analyze the LENS data with Onion clustering,\cite{becchi2024layer} an efficient algorithm for unsupervised single-point time-series clustering (see Methods for a full description of the algorithm). Briefly, hierarchically, Onion clustering identifies patterns in the time-series signals that persist longer than the time-resolution $\Delta t$, classifying them into robust dynamic micro-states, from the most populated and dominant one to the least populated, hidden ones. The method performs the clustering in a range of $\Delta t$ values (at all possible time-resolutions), from 200~ps (twice the sampling interval, the maximum time-resolution in the Onion approach) down to the minimum time-resolution (that equals to the entire trajectory length T). The method then outputs an Onion plot, which reveals the number of statistically-relevant distinct micro-clusters that can be effectively resolved as a function of the time-resolution($\Delta t$) used in the analysis (Fig.~\ref{fig:fig1}D: black line). LENS signals fluctuating faster than the resolution $\Delta t$ are labeled as unclassifiable (Fig.~\ref{fig:fig1}D, violet data: fraction of data points that do not belong to a persistent micro-environment at a certain resolution). 

The plot of~Fig.~\ref{fig:fig1}D shows how the analysis efficiently distinguishes three distinct environments (solid ice, liquid water, and the solid-liquid interface) as statistically-different micro-environments down to a resolution of $\Delta t < 1$~ns. Furthermore, at a resolution between $1 \lesssim \Delta t \lesssim 1.5$~ns the ice-water interface is decoupled in two separated/distinct environments, for a total of four environments. Fig.~\ref{fig:fig1}E shows how these correspond to molecules in bulk ice (in gray) and in bulk water phases (yellow), water molecules in contact between the solid ice (in pink), more static, and those between the interface and water, more dynamic (in green). The populations in terms of water molecules of the various micro-clusters are available in the Supporting Information (Fig. S1), while the dynamical exchanges between them are clearly visible in Movie S1. 

Recently, we demonstrated how discovering more statistically-relevant clusters in the data means "making order" and extracting information out of them, and developed an efficient method to quantify the maximum information effectively extractable from data by using a Shannon entropy minimization criterion: the Maximum Information Extraction method, MInE. For details on the method, we refer the interested readers to the dedicated paper.\cite{becchi2025maximum} We shall directly comment the results obtained when analyzing the LENS (our descriptor, $D$, in this case) time-series with the MInE method. Fig.~\ref{fig:fig1}F shows (black solid line) the information $I_{clust}$ that can be effectively extracted from the data as a function of the time resolution $\Delta t$ used to analyze the data. Briefly, such $I_{clust}$ is the sum of the $I_0$ (gray dotted line), which is the information extractable from the distribution of the data alone, plus the information gain $\Delta I$ attainable thanks to clustering of the data into distinct and statistically-relevant micro-domains (gray area).

% REVISION 
Here in this example we use Onion clustering as an efficient unsupervised clustering method capable of resolving dynamically-diverse micro-environments in the data/system, including sparsely populated ones.\cite{becchi2024layer} However, asa said above, it is worth noting how the MInE method is not restricted to one specific clustering method (such as e.g. Onion clustering \cite{becchi2025maximum}) and it is applicable to any type of data clustering (e.g., k-means, GMM, etc.).\cite{liao2005clustering,baldo2026microinformation}
The plot of Fig.~\ref{fig:fig1}F shows how $I_{clust}$ is maximum down to resolution of $\Delta t \sim 1-2$~ns -- based on this method, in such a case the optimal resolution corresponds to the lowest one guaranteeing maximum $I_{clust}$: $\Delta t \sim 2$~ns: resolution at which the number of resolved clusters is maximum (four). For lower resolutions, the $I_{clust}$ starts dropping as the fraction of unclassifiable points increases. Note that in this plot, the information ranges between 0 and 1 and is normalized in terms of how much of the data can be converted into information ($I_{clust}$) $vs.$ how much is entropy (sum of the topmost colored domains between $I_{clust}$ and $I=1$, representing the relative entropies of the various micro-environments). In Fig.~\ref{fig:fig1}F, $I_{clust}$ is thus represented in terms of fraction of data that can be converted into information.
Nonetheless, note that these plots are quantitative: $\Delta I$ can be also quantified in bits (see e.g. Fig.~\ref{fig:fig2} C, right axis), and from the differences in the entropies of the various environments (calculated at the resolution that guarantees maximum information extraction), it is possible to estimate thermodynamic entropy differences (e.g., between the ice and water environments) consistent with experimental data (e.g., with ice-water melting/freezing experimental $\Delta S$: interested readers can refer to the dedicated MInE paper for details\cite{becchi2025maximum}). 

\subsection*{The information contained in the time-derivatives}

While the approach described above allows quantifying the amount of information contained in the LENS ($D$) time-series, the same can be done, e.g., for the first time-derivative $\dot D$ ($\dot{LENS}$). In particular, it is possible to estimate the information extractable from the $\dot D$ time-series, as well as to quantify the information gain $\Delta I$ that is effectively attained when combining $D$ and $\dot D$ in a bi-variate $(D, \dot D)$ time-series analysis ($(LENS, \dot{LENS})$, in this case). 
The results obtained from the analysis of the $\dot{LENS}$ time-series (alone) are reported in the SI (Figure S2). Onion clustering resolves three separated clusters: one centered in $\dot{LENS}=0$, containing the noises of all dynamically-distinct micro-clusters (already identified by the $LENS$ analysis in Fig.~\ref{fig:fig1}) plus two symmetrical ones at positive and negative $\dot{LENS}$ values. These new clusters contain respectively melting ($\dot{LENS}>0$) and freezing ($\dot{LENS}<0$) transitions, which are effectively resolved as distinct clusters in the $\dot{LENS}$ dimension.

\begin{figure}[ht!]
 \centering
 \includegraphics[width=\textwidth,keepaspectratio]{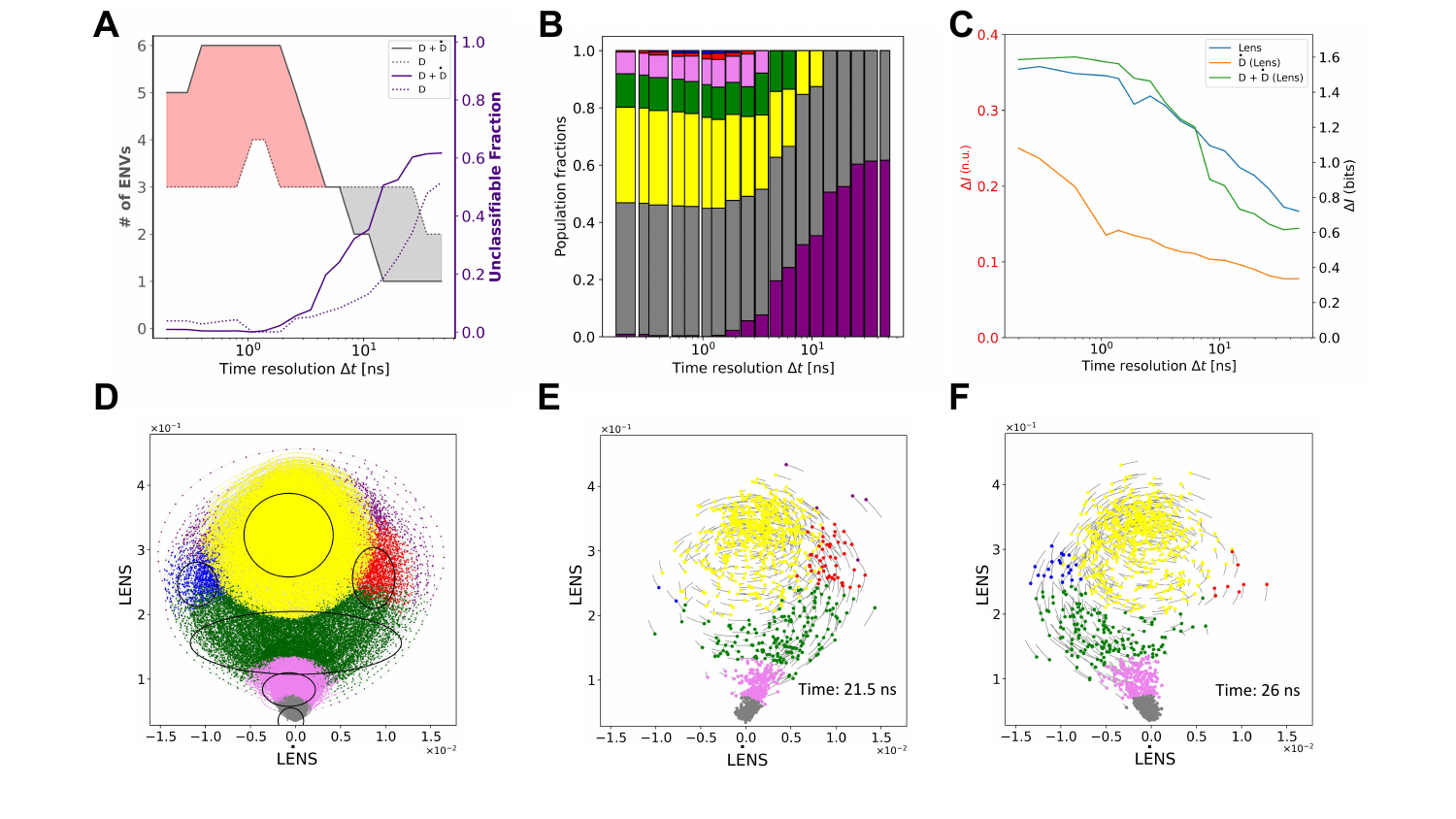}
 \caption{\textbf{Clustering analysis in a bi-dimensional ($D,\dot D$) TiDe space.} 
 (A) Comparison of Onion clustering results obtained analyzing the monovariate LENS time-series data only ($D$, dotted lines) vs. the bivariate ($LENS, \dot{LENS}$) ones (solid lines). The number of resolved micro-clusters and the fraction of unclassifiable data points as a function of the Onion clustering time resolution $\Delta t$ are indicated in gray and violet (primary and secondary Y axes, respectively). The pink area identifies the region (time-resolutions) in which the addition of the $\dot D$ in the analysis leads to a higher number of resolved environments (additional information gain). 
 (B) Population fractions of each environment as a function of the Onion clustering time resolution $\Delta t$. 
 (C) Total information gain, $\Delta I$  obtained in the analysis of the $LENS$ (blue), $\dot{LENS}$ (orange), and ($LENS$,$\dot{LENS}$) (green) time-series data as a function of the time-resolution $\Delta t$. $\Delta I$ is reported in normalized units (n.u., primary red axis) and in bits (secondary black axis).
 (D) Representation of the trajectories of all molecules in the system in the $(LENS, \dot{LENS})$ plane, each point corresponding to a certain molecule in a certain frame. Points are colored according to the clustering obtained with time resolution of e.g. $\Delta t = 0.4$~ns. 
 (E,F) Position of all the molecules in the $(LENS, \dot{LENS})$ plane at two different frames corresponding to $t=21.5$~ns (E) and $t=26$~ns (F) of MD simulation. Points are colored according to the clustering of belongings obtained from the bivariate $(LENS, \dot{LENS})$ analysis (at the time resolution of e.g. $\Delta t = 0.4$~ns). The gray tail of each point indicates the position of that same point 2 MD frames before the selected one, showing the time evolution of the molecules in the $(LENS, \dot{LENS})$ space. }
 \label{fig:fig2}
\end{figure}

Such two additional clusters are also captured when analyzing in the same way the bivariate $(D, \dot{D})$ time-series. The Onion plot of Fig.~\ref{fig:fig2}A shows that the number of detected clusters in the bivariate analysis (solid lines) clusters resolved by Onion clustering increases from the four detected in conditions of maximum information extraction when considering the $LENS$ data alone (dotted data, reported for comparison) up to six. 
The detection of such six micro-domains is stable down to a resolution of $\Delta t \sim 2$ ns. This demonstrates how adding the $\dot{LENS}$ dimension effectively adds information to the analysis: namely, it allows extracting both the information contained in the four clusters resolved in the $LENS$ data and the information contained in the molecular transitions between them (contained in the $\dot{LENS}$ data).
Fig.~\ref{fig:fig2}B shows the population of the clusters resolved as a function of the time-resolution of the analysis. Although in this system the population of these two additional transition $\dot{LENS}$ clusters is relatively small (blue and red) compared to those of the four $LENS$ clusters, these nonetheless provide additional information that can be quantified. 
Fig.~\ref{fig:fig2}C plots the information extracted during the analysis of the $LENS$ (in blue), $\dot{LENS}$ (orange), and ($LENS$,$\dot{LENS}$) (green) time-series data as a function of the time-resolution $\Delta t$. Evidently, large part of the information contained in the $\dot{LENS}$ (orange) is redundant to that of $LENS$ (in particular, that contained in the main $\dot{LENS}=0$ peak environment: see Fig. S2). However, quantifiable information gain (Fig.~\ref{fig:fig2}C: green $vs.$ blue data) is evident for resolutions $\Delta t \lesssim 4$ ns, in the same regime in which more environments can be effectively resolved in the bivariate ($LENS$,$\dot{LENS}$) time-series as compared to the $LENS$ data alone (Fig.~\ref{fig:fig2}A: pink area identifying positive information gain). Below this resolution, the inclusion of $\dot{D}$ instead reduces the efficiency of the analysis, resulting in information loss (Fig.~\ref{fig:fig2}C: green $vs.$ blue data): as seen in many other cases recently, the number of environments detected decreases and the fraction of unclassified data points increases compared to analyzing the monovariate $D$ analysis (Fig.~\ref{fig:fig2}A: gray area) due to noise-to-noise additions and information frustration phenomena.\cite{lionello2025relevant,becchi2025maximum} 

While the gray, pink, green, and yellow clusters detectable along the $LENS$ dimensions (Fig.~\ref{fig:fig2}D-F: $y$ axis) identify microscopic domains differing from each other in terms of the average $LENS$ (local reshuffling dynamics) of the molecules populating them (within the same cluster, molecules have similar local dynamics), the red and blue clusters spottable along the $\dot{LENS}$ dimension (Fig.~\ref{fig:fig2}D-F: $x$ axis) identify particles that, over a consistent time interval, increase or decrease their LENS signal in similar ways (consistent rates of change of $LENS$ in $\delta t$): these thus indicate that there are patterns in the melting and freezing transitions hidden in the data that are gathered while then clustering in the $\dot{LENS}$ dimension.
At the molecular level, the two new $\dot{D}$ transition environments are located near the solid-liquid interface (see supplementary Movies S2 and S3). 
These represent dynamically active regions that are crucial for the physics and dynamical phase interactions in the system, and that can be captured only in the $\dot{D}$ dimension but remain otherwise hidden while navigating the $D$ dimension alone (Fig.~\ref{fig:fig2}D: blue and red clusters can be discriminated in a 2D $LENS$,$\dot{LENS}$ plot only along the $\dot{LENS}$ axis). 
Fig.~\ref{fig:fig2}E-F show two snapshots taken at different times during the MD simulation. The oxygen atoms of water molecules are colored based on the clusters they belong to in these frames. Their small linear tails connect each point to the point they were in the $LENS$,$\dot{LENS}$ plane two stored frames earlier: this provides a clear insight into how molecules in transient states evolve over time. 
Furthermore, these plots also show that red and blue transitions are not always present and have oscillatory character over time (Fig.~\ref{fig:fig2}E shows a large number of red melting-transition
molecules and very few blue freezing-transition molecules, whereas
Fig.~\ref{fig:fig2}F shows the opposite trend). 

\begin{figure}[ht!]
 \centering
 \includegraphics[width=\textwidth,keepaspectratio]{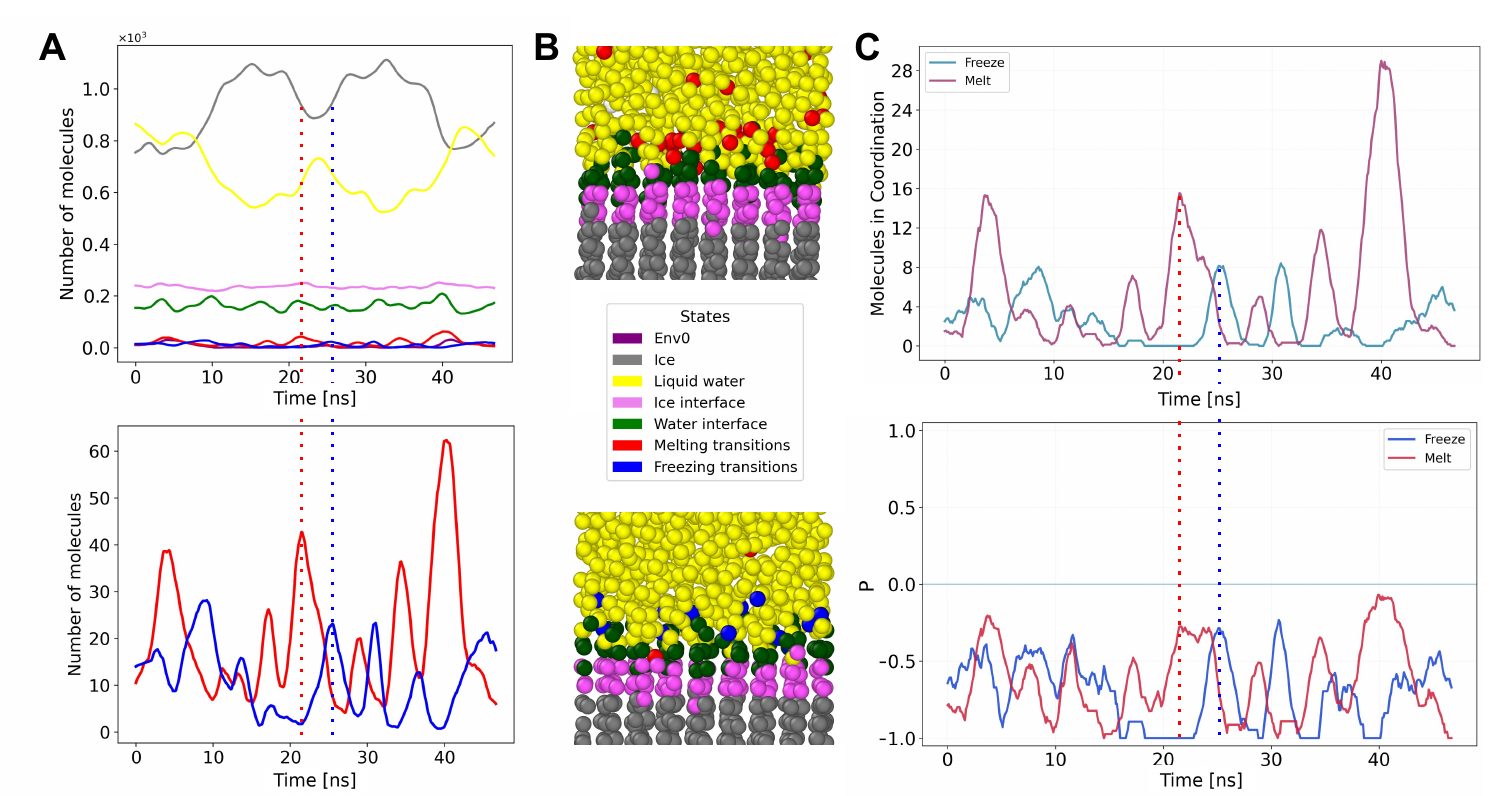}
 \caption{\textbf{Physical-chemical interpretability of the information gained in the dimensions of a TiDe space.} (A) Top: population of molecules assigned to each cluster as a function of time, obtained via Onion clustering of the $(LENS, \dot{LENS})$ timeseries at the time resolution of e.g. $\Delta t = 0.4~\mathrm{ns}$ (micro-clusters legend on the right). Bottom: population of molecules assigned to the ``melting-transition'' and ``freezing-transition'' (red and blue micro-clusters -- discernible in the $\dot{LENS}$ dimension --, respectively) as a function of time. The two vertical dotted lines mark two representative frames corresponding to peaks in melting and freezing transitions, respectively. (B) Representative snapshots of the MD trajectory at the two frames highlighted in panel (A): top melting-transition cluster at $t=21.5~\mathrm{ns}$; bottom, freezing-transition cluster at $t=26.0~\mathrm{ns}$ (color code as in the legend). (C) Top: number of molecules undergoing collective melting or freezing transitions (i.e., where molecules melt, or freeze, simultaneously as spatially-correlated molecular clusters) as a function of MD simulation time, in light red and light blue, respectively. Bottom: parameter $P=\frac{C-S}{C+S}$, where $C$ is the number of molecules in spatially connected transition clusters and $S$ is the number of molecules undergoing sparse individual transitions, ranges from -1, for a dynamics purely dictated by sparse single-molecule de-correlated melting (red) or freezing (blue) transitions, to 1, for purely collective internal dynamical transitions (molecules undergoing transitions only in clusters). $P=0$ means that both local or collective dynamical transitions have similar statistical weight. The data show that local transitions are dominant in this system, although individual collective melting/freezing events are also observed (see e.g. the two vertical dotted lines corresponding to the same MD frames highlighted in panels (A) and (B), where two collective melting and freezing transitions are observed).}
 \label{fig:fig3}
\end{figure}

This brings us directly into the facile interpretability and usefulness of the information contained in the different TiDe dimensions.
As can be observed in Fig.~\ref{fig:fig3}A (top panel), the dynamic complexity of the system is hierarchical and multilayered.
The populations of the clusters (in terms of number of molecules) oscillate around equilibrium values. In particular, while those associated with the ice-water interface environments (green and pink curves) exhibit stable molecular populations over time, those of the two main ice and liquid clusters (respectively, in gray and yellow) fluctuate considerably, with oscillation amplitudes up to $\sim 200-250$ molecules, on long time-scales ($\sim 10^1$ ns).
Looking more closely at the populations of the red and blue transition clusters (Fig.~\ref{fig:fig3}A, bottom), it is clear that melting and freezing transitions also fluctuate over time.
The two vertical dotted red and blue lines in Fig.~\ref{fig:fig3}A identify two example frames at which the melting and freezing transition clusters are highly populated, for which representative snapshots of the system are shown in Fig.~\ref{fig:fig3}B, where at $t=21.5$~ns of simulation (top) the system is rich in melting molecules (in red: molecules undergoing transition in the same time-interval toward a more liquid/dynamical state) while at $t=26$~ns (bottom) the same happens for freezing molecules (in blue).
The vertical red dotted line in Fig.~\ref{fig:fig3}A identifies a frame where the drop in the ice population (gray) and the rate of growth of the liquid phase (yellow) are maximum: in that time-interval the number of molecules undergoing melting transition has a peak (Fig.~\ref{fig:fig3}A, bottom: in red) while there are almost no freezing molecules (blue curve close 0). Conversely, the vertical blue dotted line (Fig.~\ref{fig:fig3}A) identifies a successive time-frame along the simulation where the rate of growth of the ice phase and the rate of drop of the liquid phase (yellow) are maximum, the number of molecules undergoing freezing transitions have a peak while the melting one are at the minimum (Fig.~\ref{fig:fig3}A, bottom: in blue and red respectively). 

The plot of Fig.~\ref{fig:fig3}A (bottom) thus shows how both melting and freezing transitions are not uniform in time, but tend to appear as coordinated (simultaneous) in time.
The plots of Fig.~\ref{fig:fig3}C provide additional insights into the collective nature of such melting and freezing transitions. For each of these transitions over time, we can discriminate how many of these molecules undergo transitions as spatially coordinated/correlated (collective transitions) or as decorrelated in space (individual local transitions). Fig.~\ref{fig:fig3}C (top) shows the number of molecules undergoing collective melting (pink) and freezing (cyan) transitions. This is evaluated by calculating how many transitions occur within the first solvation shell of each transiting molecule. In the time-frame identified by the red vertical dotted line, when a total of $\sim 40$ molecules are melting at the same time (Fig.~\ref{fig:fig3}A, bottom), $\sim 16$ of them are also undergoing transitions in spatially coordinated way (Fig.~\ref{fig:fig3}C, top: in pink, collective transitions) while the remaining $\sim 24$ are thus sparse local transitions that are not correlated in space. In correspondence of the blue dotted vertical line,  when a blue peak of $\sim 22$ freezing molecules is observed in Fig.~\ref{fig:fig3}A (bottom), the cyan plot in Fig.~\ref{fig:fig3}C (top) shows that $\sim 8$ of them are freezing as coordinated in space (collective freezing transition) while the remaining $\sim 14$ are sparse local freezing transitions not coordinated in space.

From these data, we quantify the local versus collective character of the transitions frame by frame using the parameter
\begin{equation}
P=\frac{C-S}{C+S},
\end{equation}
where $C$ is the number of molecules undergoing transitions in a spatially connected/clustered way, while $S$ is the number of molecules undergoing sparse, spatially decorrelated transitions. The extreme values $P=1$ and $P=-1$ identify frames dominated respectively by collective or sparse/local transitions, while $P=0$ indicates equal collective and sparse contributions.
    
The trend in Fig.~\ref{fig:fig3}C (bottom panel) shows that, throughout the entire simulation, the value of $P$ is always $<0$. This indicates how in this system local transitions are always prominent as compared to collective ones, which, nonetheless, reach comparable weight to local ones in specific time-frames that emerge every $\sim 5-10$ ns. This reveals the complex multilayered internal dynamics of such a system, where sparse transitions occur continuously over time (see also Fig. S3), which are accompanied by "waves" of collective transitions that allow the main phases of the system (ice and liquid water) to breathe growing and shrinking over time.

It is worth noting that such deep insights into the internal physics of the system are possible and facilitated in this case thanks to the fact that red and blue clusters are readily resolved along the $\dot{LENS}$ dimension.  
This demonstrates the utility of the TiDe formalism, which allows gathering different types of complementary information in the different dimensions. This concept, which has been here demonstrated by navigating the TiDe space up to the second dimension ($D$ and $\dot D$), can be generalized up to the $n^{th}$ TiDe dimension ($D^{(n)}$).

\begin{figure}[ht!]
 \centering
 \includegraphics[width=\textwidth,keepaspectratio]{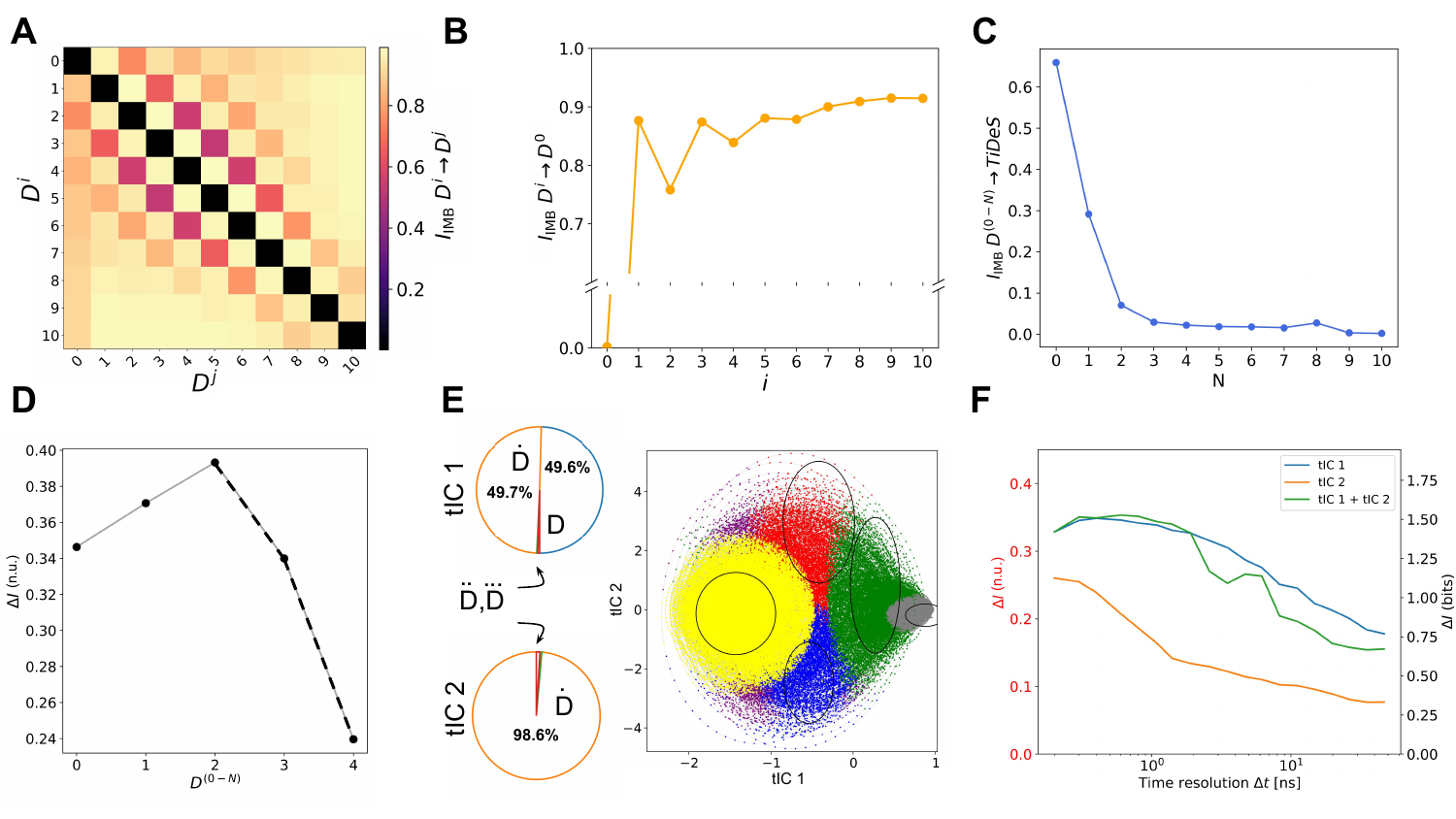}
 \caption{\textbf{Complementarity of the information contained in the TiDe dimensions.}
 (A) Information Imbalance matrix $I_{\mathrm{IMB}}(D^i\rightarrow D^{j})$ between the TiDe components from $D$ to $D^{(10)}$, ranking the similarity in the information content between the various components. $I_{\mathrm{IMB}}(D^i\rightarrow D^{j}) =0$ indicates that the two TiDe dimensions contain the same information, while $I_{\mathrm{IMB}}(D^i\rightarrow D^{j})=1$ that these contain different information.\cite{glielmo2022ranking} The analysis shows strong imbalance $I_{\mathrm{IMB}}(D^i\rightarrow D^{j})>>0$ between consecutive-order time-derivatives and lower values when skipping one time-derivative order, consistent with an oscillating signal structure (e.g., as in an ideal sinusoidal signal). 
 (B) Information Imbalance of each derivative order with respect to the original descriptor, $I_{\mathrm{IMB}}(D^i\rightarrow D^0)$, highlighting strong imbalance variations for the first derivative orders followed by a plateau at higher orders.
 (C) $I_{\mathrm{IMB}}$ between growing TiDe subsets $\mathcal{D}^{(0-N)}=\left(D^0,D^{(1)},\dots,D^{(N)}\right)$ and the full TiDe representation, showing how progressively adding derivative coordinates improves the reconstruction of the full TiDe information. 
 (D) Information gain, $\Delta I$ obtained with Onion clustering as a function of the maximum derivative order $N$ included in the TiDe subset $\mathcal{D}^{(0-N)}$. The solid line indicates the regime where adding derivative coordinates increases the information gain, while the dotted line indicates the regime where additional derivative coordinates degrade clustering performance. (E,F) tICA time-series analyses. (E) Left: composition of the tIC1 and tIC2 components in terms of content of information coming from the original TiDe dimensions. Right: representation of the trajectories of the molecules in the $(\mathrm{tIC1},\mathrm{tIC2})$ plane, colored according to Onion clustering at time resolution $\Delta t = 0.4$~ns (these tICA analyses loose one interface environment).
 (F) MInE calculated $\Delta I$, \cite{becchi2025maximum} obtained with Onion clustering as a function of the time resolution $\Delta t$ by analyzing the monovatiate tIC1 (blue), tIC2 time-series data (orange), vs. the bivariate (tIC1,tIC2) time-series (green) data: this comparison shows that the addition of tIC2 provides little or no additional information gain in this case, reaching a maximum information gain lower than that of Fig. \ref{fig:fig2}C (i.e., without using any TiDe dimensionality reduction); $\Delta I$ is reported in normalized units (n.u., primary red axis) and in bits (secondary black axis).}
 \label{fig:fig4}
\end{figure}

\subsection*{Proving and quantifying information complementarity}

While we demonstrated how adding the first derivative $\dot D$ substantially enhances the amount of information extracted during the analysis, it is interesting to assess the information contained in the higher-order derivatives, and to unveil the relative contribution that these may provide to the analysis. To this end, we constructed a comprehensive 11-dimensional TiDe containing all derivatives time-series data $D(t)$ (base data, $D^0$: 1$^{st}$ TiDe dimension), $\dot D(t)$ (1$^{st}$ time-derivative, 2$^{nd}$ TiDe dimension), $\ddot D(t)$, ..., to $D^{10}(t)$ (see also \ref{eq:eq2}).
To complement the results of Fig.~\ref{fig:fig2}, and in order to assess the amount and mutual complementarity of the information contained in the individual TiDe dimension, we analyzed the data with the Information Imbalance framework~\cite{glielmo2022ranking}. This provides a quantitative comparison of the information contained in the data contained in the various time-derivatives dimensions of the TiDe space, capturing redundancy and complementarity in their informational content (see Methods for details).

The matrix reported in Fig.~\ref{fig:fig4}A shows the mutual info imbalance ($I_{\mathrm{IMB}}$) values between the different order time-derivatives with each other: $I_{\mathrm{IMB}}(D^i \rightarrow D^j)$, with $0 \leq i,j \leq n$ (and $n=10$ in the specific example discussed herein).
The various $i,j$ cells in the matrix are colored based on the $I_{\mathrm{IMB}}(D^i \rightarrow D^j)$ values, with darker color being lower $I_{\mathrm{IMB}}$. 
Black cells identify $I_{\mathrm{IMB}}=0$ values, meaning that the two components contain exactly the same information (Fig.~\ref{fig:fig4}A: $I_{\mathrm{IMB}}(D^j \rightarrow D^j)=0$ by definition). A hypothetical $I_{\mathrm{IMB}}=1$ means that the two components are completely orthogonal (they contain completely different information).\cite{glielmo2022ranking}

A considerably high information imbalance $I_{\mathrm{IMB}}$ is observed between consecutive derivatives: in particular, $I_{\mathrm{IMB}}(D \rightarrow \dot D) \gtrsim 0.9$, $I_{\mathrm{IMB}}(\dot D\rightarrow \ddot D) \gtrsim 0.9$, etc. (Fig.~\ref{fig:fig4}A), while time-derivatives differing by two orders contain information that is more similar to each other (e.g., $I_{\mathrm{IMB}}(\dot D \rightarrow \dddot D)$, $I_{\mathrm{IMB}}(\ddot D \rightarrow D^{4})$). 
Such a pattern indicates that the signals captured by the derivatives follow an oscillatory behavior, where consecutive TiDe dimensions contain complementary information, while deriving them again restores similarity compared to the original information (similar to an ideal hypothetical case where the signal is a function like $D(t)=\sin t$ and one thus has$\dot D(t)=\cos t$, $\ddot D(t)=-\sin t$, etc., periodically recovering the same information).
Fig.~\ref{fig:fig4}B reports the imbalance between all different-order derivative (each TiDe dimension) and the initial LENS $D$ signal ($I_{\mathrm{IMB}}(D^i \rightarrow D^0)$). This shows a sharp increase for $I_{\mathrm{IMB}}(\dot{D} \rightarrow D) \sim 0.9$, followed by a plateau at higher orders. The high value of
$I_{\mathrm{IMB}}(\dot D\rightarrow D)$ indicates little correspondence
between the neighborhood structures defined by $\dot D$ and $D$.

Combining together a variable number of TiDe dimensions, it is possible to assess how much of the information contained in the entire TiDe space can be effectively recovered (i.e., all 11 time-derivatives $D$ to $D^{10}$ taken together -- we indicate it as: ($D^{0-10}$).
The individual TiDe space dimensions that is most descriptive of the system is the one containing the original $LENS$ data ($D$) that, as shown in the plot of Fig.~\ref{fig:fig4}C, allows on its own to reach  $I_{\mathrm{IMB}}(D^0 \rightarrow D^{0-10}) \sim 0.65$.
The most descriptive couple of TiDe components are the two first time-derivatives taken together, which produce a drop of $I_{\mathrm{IMB}}\!\left((D,\dot D)\rightarrow\mathcal{D}^{(0-10)}\right)\sim0.29$.
Plateau at $I_{\mathrm{IMB}} \sim 0$ is reached when adding beyond the 3$^{rd}$ TiDe dimension to the analysis.

This means that an additional information gain in the analysis is obtained when the $\ddot D$ data are also considered together with $D$ and $\dot D$.
However, the $I_{\mathrm{IMB}}$ analysis indicates that adding additional time-derivatives and going higher-dimensional does not produce any relevant additional information gain.

This was also proven with another independent method for information extraction from time-series data - MInE, which essentially measures the drop and minimization of Shannon entropy following to discovering robust clusters in the data.\cite{becchi2025maximum}
We analyzed with Onion clustering mono- to penta-variate time-series data composed of the ($D$), ($D,\dot D$), ($D,\dot D,\ddot D$),  ($D,\dot D,\ddot D,\dddot D$),($D,\dot D,\ddot D,\dddot D,\ddddot D$) and systematically estimated with the MInE method the amount of information extractable from the structure/patterns in the time-series data. 
Fig.~\ref{fig:fig4}D shows how -- consistently with the $I_{\mathrm{IMB}}$ analysis results -- the information gain ($\Delta I$) increases while adding up to the $\ddot D$ derivative, while adding more produces a drop due to the addition of useless components which contain no additional information but, nonetheless, non-negligible noise (analysis of larger-dimensional time-series containing more than four time-derivatives dimensions produces worse-and-worse results).
In particular, the Onion clustering and MInE analysis indicate how the information gain (Fig.~\ref{fig:fig4}D: $\Delta I$) obtained following to the addition of the $\dot{LENS}$ data to the $LENS$ ones is related to the discovery of the two new physical clusters in the TiDe space: the red and blue transition clusters of Figs.~\ref{fig:fig2},~\ref{fig:fig3}. 
The additional $\Delta I$ that is attained while adding also the $\ddot{LENS}$ data are instead not due to the discovery of an additional separate cluster (which remains constant when moving from a 2D to a 3D time-series analysis), but rather to a decrease in the residual unclassifiable points (ENV0) and to a better and more complete classification of the data-points in the various clusters that are discovered in the TiDe.
Taken together, both information imbalance and MInE analyses indicate that the $LENS$ data analyzed for this ice-water system are in reality 3-dimensional in the TiDe space: namely, there is information nested in up to 3 dimensions.

\subsection*{Information interpretability without dimensionality reduction}

When dealing with high-dimensional datasets, dimensionality reduction is typically a necessary step to obtain physically-interpretable results.
One key advantage of the TiDe space formalism, is the direct interpretability of the information contained in the various TiDe dimensions, which does not need dimensionality reduction. 
This is a considerable difference with, e.g., a time-delayed embedding (Takens's embedding theorem).\cite{noakes1991takens,takens2006detecting} In fact, while this allows to guarantee maximum information extraction as when accounting for all time-derivatives in a TiDe space analysis, in the case of a time-delayed embedding it is not as straightforward as in TiDe spaces to reconduct the different bits of information to different types of (orthogonal and complementary) physical events/phenomena. 
Such interpretability is on the other hand intrinsic in the complementarity and orthogonality of the different TiDe space's dimensions.

As an illustrative proof-of-concept example, we analyze the entire 11-dimensional $LENS$ time-series TiDe space analyzed in the previous section with a Time-lagged Independent Component Analysis (tICA)~\cite{molgedey1994separation, naritomi2013slow}. 
tICA is a dimensionality-reduction approach that is widely used to analyze time-series data: it allows to identify the slowest de-correlating modes in time-series data, thereby capturing, in principle, all essential dynamical processes with a minimal set of collective variables (see Methods for details).

The test aimed at verifying if a dimensionality reduction such as tICA was capable of reducing the dimensionality of the 11D TiDe dataset to the 2-3 main dimensions seen as the necessary ones to capture all key dynamical features of the system in Fig.~\ref{fig:fig4}A-D (or to very similar ones).
Fig.~\ref{fig:fig4}E left shows the two main tIC1 and tIC2 components after tICA dimensionality reduction, which are indeed mainly composed of information contained in the first two TiDe dimensions ($D$ and $\dot D$), with negligible contributions from higher-orders time-derivatives (and no contribution at all from time-derivatives above the third-order one $\dddot D$). 
On the one hand, this provides another independent confirmation that derivatives above the second order add little meaningful information to the dynamics.
On the other hand, the results of Fig.~\ref{fig:fig4}E (left) demonstrate how, while the tIC2 component is composed of $\sim98-99 \% $ of information contained in the $\dot D$ ($\dot{LENS}$) time-series (2$^{nd}$ TiDe dimension), the content of tIC1 is more ambiguous, and is $\sim 50-50 \%$ composed of information contained in both $D$ and $\dot D$ TiDe dimensions.
This means that tICA is not capable of well-discriminating the different dynamical nature of the physical events which data are contained in the $LENS$ and $\dot{LENS}$ time-series data.
We also performed mono- and bi-variate time-series analysis of the tIC1 and tIC2 data, as done with $LENS$ and $\dot{LENS}$ in Fig.~\ref{fig:fig2}.
The clustering results projected in the (tIC1,tIC2) plane in Fig.~\ref{fig:fig4}E (right) show a picture that is similar, but less well-resolved than without any dimensionality reduction. Compared to Fig.~\ref{fig:fig2}D, the analysis loses one cluster (one of the two sub-interface environments) and, most importantly, there is also less neat division between the LENS environments and the transition between them. Note how the red and blue transition clusters "touch each other" at $tIC2=0$ and, when all clusters are projected on either the tIC1 or tIC2 dimensions, are way more superimposed to green and yellow environments as compared as in Fig.~\ref{fig:fig2}D.

We used again the MInE approach to quantify the Information gain extractable from the tIC1, tIC2, and (tIC1,tIC2) time-series as a function of the clusters effectively resolvable within them at the various time-resolutions ($\Delta t$). 
The results in Fig.~\ref{fig:fig4}F show that, while combining tIC1 provides an information comparable and slightly lower than that extractable from the $LENS$ time-series alone (Fig.~\ref{fig:fig2}C, in blue), in this case adding also tIC2 in a bi-variate time-series analysis does not yield any appreciable information gain (Fig.~\ref{fig:fig4}F: green curves lower or at best superimposed to the blue curve). This demonstrates how all the information is in this case already concentrated in the tIC1 dimension.
In the tICA analysis the problem thus appears to be mono-dimensional, while from the results of Fig.~\ref{fig:fig3} and Fig.~\ref{fig:fig4}A-D it is clear that it is 3-dimensional.
All this evidence indicates that the dimensionality reduction (even a tICA one, typically considered as well-suited for time-series data) is not only not necessary to analyze a TiDe space, but it can be also detrimental essentially for two reasons: 1) it does not provide any information gain, 2) it also compromises physical interpretability by mixing information related to different physical event types (e.g., $D$ and $\dot D$) into individual dimensions (e.g., tIC1) and making it difficult to decouple them \textit{a posteriori}.

In this sense, it is worth underlining that tICA, used in the comparison discussed above as a reference example, is based on a linear decomposition, which could be at the origin, to some extent, of the Fig.~\ref{fig:fig4}E,F. Other methods can eventually be used to this end, which can potentially reduce and mitigate such distortions, such as, e.g., the variational approach for Markov processes (VAMP)\cite{wu2020vamp} -- VAMPnets implement this approach through neural networks, learning nonlinear features and probabilities of belonging to different states\cite{mardt2018vampnets}--, or, e.g., the State Predictive Information Bottleneck (SPIB) -- which learns a compressed representation useful for predicting future state labels and refines those labels during training.\cite{wang2021spib}
While a systematic testing and comparison between all such methods is out of the scope of this paper, it is worth underlining that the comparison of Fig.~\ref{fig:fig4} is interesting in that it provides a clear example of how, \textit{per se},  TiDes analysis do not require dimensionality reduction \textit{a priori} to provide interpretable results. The fact that each single TiDe dimension contains information related to physically different events makes it interesting to keep them separated rather than compressing them \textit{via} dimensionality reduction, with no compromise on the interpretability of the results. 

%\textcolor{blue}{Once informative coordinates are available, they can also serve as inputs to methods that learn compact representations of system evolution. These methods seek combinations of the input variables that retain dynamical relationships between observations separated in time. Depending on the chosen time interval, this can emphasize persistent environments and exchanges between them over rapid fluctuations. The variational approach for Markov processes (VAMP) provides a framework for identifying such combinations, including for nonreversible processes.\cite{wu2020vamp} VAMPnets implement this approach through neural networks, learning nonlinear features and probabilities of belonging to different states.\cite{mardt2018vampnets} A related goal is pursued by the State Predictive Information Bottleneck (SPIB), which learns a compressed representation useful for predicting future state labels and refines those labels during training. Its prediction delay controls which dynamical distinctions remain resolved.\cite{wang2021spib}

\subsection*{Example 2: TiDe exploration from experimental tracking of active colloidal systems}

\begin{figure}[ht!]
 \centering
 \includegraphics[width=\textwidth,keepaspectratio]{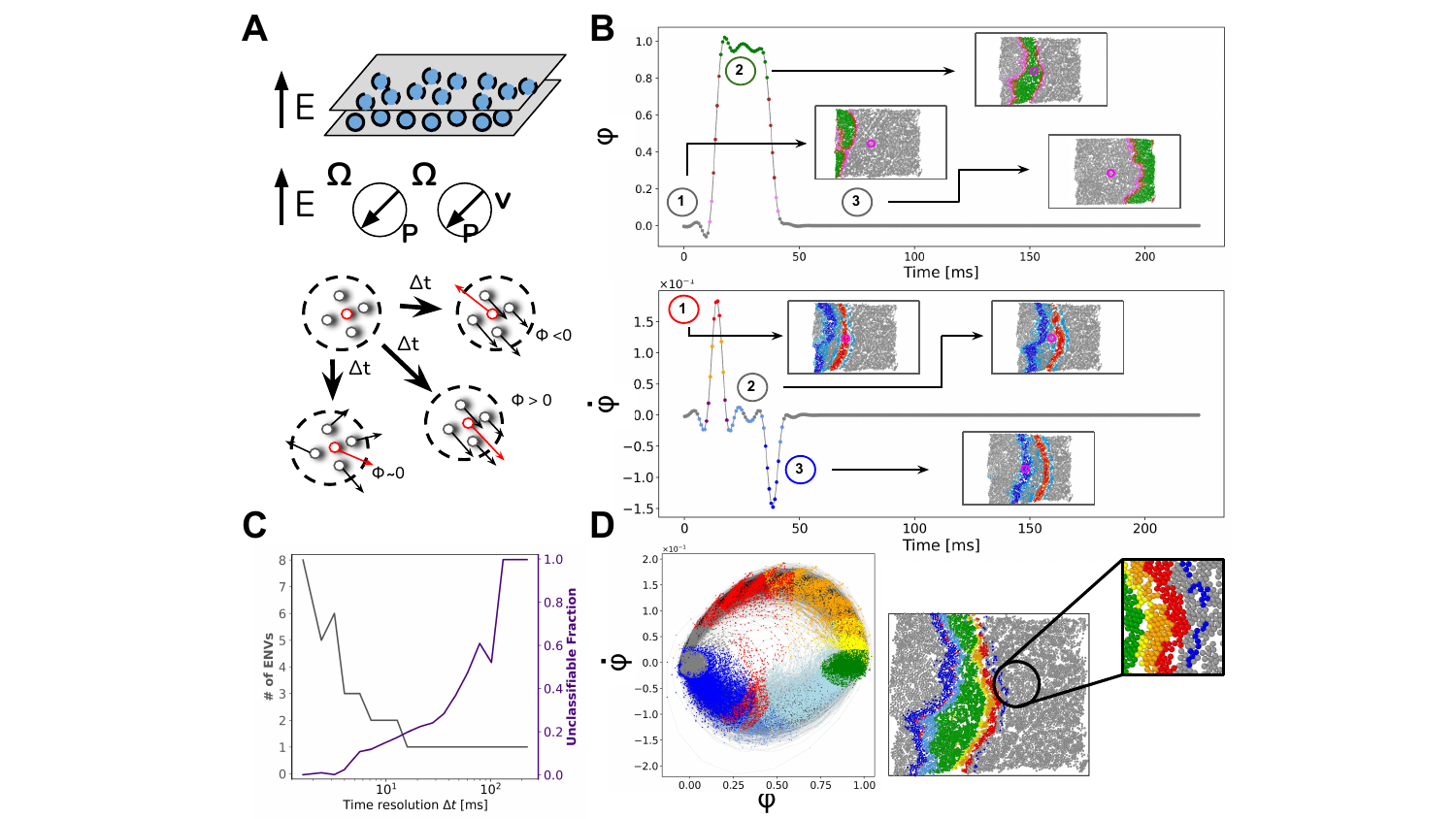}
 \caption{
 \textbf{Navigating TiDe spaces for higher-scale experimentally-tracked active matter complex systems.}  
 (A) Top: cartoon representation of Quincke rollers, colloidal polystyrene micro-particles with a residual internal dipole moment confined in the $xy$ plane that, when immersed in a conducting fluid, display complex collective dynamics under a weak DC electric field applied orthogonally to the plane.\cite{liu2021activity,becchi2024layer,doria2025data,lionello2025relevant,caruso_pnas_nexus_classification} Bottom: schematic illustration of the velocity alignment $\varphi$ descriptor used for this analysis, which captures collective motion, showing how $\varphi$ close to 0 indicates no alignment detected, $\varphi>0$, particles with same alignment and $\varphi<0$ anti-alignment of the particles around a center. 
 (B) Representative single-particle trajectory clustered by analyzing $\varphi$ and $\dot{\varphi}$. Top: raw descriptor $\varphi$ ($D$) showing transitions between quiescent state (1), wave core (2), and post-wave relaxation (3). Bottom: first time-derivative $\dot{\varphi}$ ($\dot{D}$) of the same trajectory, resolving alignment increase at the accelerating front (1), no alignment changes in the maximum alignment area at the wave core (2), and de-alignment as the wave goes away (3).  
 (C) Onion clustering results as a function of the time resolution $\Delta t$ for the two-dimensional dataset $(D, \dot{D})$, showing the number of detected environments (gray) and the fraction of unclassified data points (purple). Relevant environments emerge primarily at the highest accessible resolutions, reflecting the limited sampling of experimental data. 
 (D) Left: particle trajectories in the $(D,\dot{D})$ plane, colored according to Onion clustering at $\Delta t=1.6~\mathrm{ms}$. Right: representative spatial configuration showing the resolved dynamical environments. The clustering resolves up to eight distinct environments, distinguishing the wave core from accelerating and decelerating fronts and capturing the full propagation cycle of the collective wave.
 }
 \label{fig:fig5}
\end{figure}

To demonstrate the generality of the approach, we built a TiDe space from data extracted from experimentally-resolved trajectories of a mesoscopic colloidal system composed of a large number of active polymeric microparticles, called Quincke rollers~\cite{liu2021activity} (see Methods for details). We do not enter into the details of such interesting active matter systems, and we refer interested readers to the dedicated paper from which the movies of their interactions have been obtained by Liu et al.~\cite{liu2021activity}. 
In brief, such mesoscopic polystyrene microparticles (diameter $\sim10~\mu$m) move driven by a weak perpendicular electric field, under which they have been reported to give rise to complex dynamic behaviors, collective waves, vortices, etc. (Fig.~\ref{fig:fig5}A).
Recently we resolved the trajectories of all Quincke rollers particles from an experimental movie where these are observed over time at the microscope,\cite{becchi2024layer,caruso_pnas_nexus_classification,lionello2025relevant,doria2025data,martino2025dynsight} which allows us to use exactly the same analyses that can be run for a simulated system as, e.g., the aqueous one described in the previous sections.
As an additional case study, we thus here analyze the trajectories of $i = 6921$ tracked particles in a $700 \times 700$~$\mu$m$^2$ microscopy field of view, recorded over a duration of $\sim0.25$~s. The trajectories contain a total of $T=312$ frames (one every $\Delta \tau=0.8$ ms, which constitutes the base temporal resolution limit in the analysis in this case, see also below). Within this window, a collective wave of particles crosses the system, propagating from left to right (see Video S4).

As a descriptor of concerted collective motions, here we monitor the alignment of the velocities of the particles within a cutoff radius from each particle $i$ in the system,$\varphi$~\cite{vicsek2012collective}, which is an efficient descriptor for detecting collective and coordinated motions on many-body complex dynamical systems (Fig.~\ref{fig:fig5}A, see also Methods for details).\cite{becchi2024layer,lionello2025relevant,doria2025data,caruso_pnas_nexus_classification} 
As done in the ice/water example above, again we  proceeded to construct a TiDe space of excess-dimensionality and started from calculating the time-derivatives of $D(t)$ ($\varphi(t)$, in this case) up to $D^{10}(t)$. 
However, in this specific case where the analysis is implicitly limited by the experimental data acquisition frequency (one frame every $0.8$~ms), information-containing components can be practically obtained only up to the $2^{nd}$ TiDe dimension $\dot D(t)$: $\dot \varphi(t)$. Higher order time-derivatives of $\varphi(t)$, in fact, were found to be dominated by scattered points, and no information could be practically extracted from them. This indicates that this specific dataset is, in the $\varphi(t)$ TiDe space, effectively bi-dimensional (information present only in $\varphi(t)$ and $\dot \varphi(t)$).  

Again, we used the Onion clustering approach to analyze the mono- and bi-variate time-series: $\varphi$, $\dot \varphi$ and ($\varphi$,$\dot \varphi$). 
Effectively, there is a discovery of new clusters in the analysis when considering both TiDe dimensions (($\varphi$,$\dot \varphi$) -- see Fig. S4 in the SI).  
Fig.~\ref{fig:fig5}B reveals details of the left-to-right Quincke rollers' wave propagation. The top panel illustrates the $\varphi$ time-series of one representative particle in the system. This undergoes sequential transition across different states: (1) initially quiescent, (2) then swept into the propagating wave, (3) finally returning to a quiet post-wave state. The bottom panel in Fig.~\ref{fig:fig5}B shows the corresponding $\dot \varphi$ time-series for the same particle, demonstrating positive and negative $\dot \varphi$ peaks when the particle undergoes respectively $1 \rightarrow 2$ and $2 \rightarrow 3$ transitions in the top panel.

From the bi-variate $\varphi,\dot \varphi$ time-series, Onion clustering resolves up to eight distinct dynamical domains and transient micro-states/clusters. 
Fig.~\ref{fig:fig5}D shows how the eight discovered clusters correspond to different well-defined domains in the system: from quiescent areas and the wave "core" ($\varphi$ states, in gray and green respectively), to the accelerating front (shades of orange) and decelerating back (shades of blue) of the propagating wave.
The Onion plot of Fig.~\ref{fig:fig5}C shows how this is possible only at very high temporal resolutions (at the lowest $\Delta t=2$ frames), while the number of clusters effectively resolvable in the analysis decays very quickly while increasing the $\Delta t$. This confirms that in this specific case the temporal sampling of the microscopy trajectory is barely sufficient to sustain such a resolution in the analysis. 
Nonetheless, note that at least 5 clusters (corresponding to the quiescent domain, wave core, accelerating and decelerating domains, plus unclassifiable points, in the ENV0 cluster) can be stably detected down to the resolution of $\Delta t \sim 4$ frames). Lowering the resolution further implies the loss of the transition accelerating/decelerating domains as distinct clusters, which become part of the unclassifiable ENV0 cluster (for a total of 3 clusters: quiescent domain, wave core, and ENV0).

Interestingly, at coarser resolutions with $\Delta t\geq20$ frames, only one
environment is resolved, suggesting a characteristic residence time
of order 20 frames ($\sim16$~ms) for the wave-associated states.

Notably, the clock-wise sequence transition of each particle $i$ in the system across these micro-states in the $\varphi,\dot \varphi$ plane of Fig.~\ref{fig:fig5}D (left) independently verifies the wave without any \textit{a priori} assumption in terms of directionality (see also Supplementary movie S5). 
This demonstrates how not only TiDe spaces offer an ideal framework to maximize information extraction from time-series data of virtually any type of system, but also provide a reliable data-driven framework for uncovering key local and collective phenomena or events in systems about which little is known \textit{a priori} (such as e.g. the in active matter system used herein as an example) and, in that, for facilitating the study and understanding of complex dynamical systems.

\section*{Conclusions}

In this work, we demonstrate how, starting from virtually any type of time-series data, it is possible to build a high-dimensional Time-Derivatives (TiDe) space where the individual TiDe's dimensions are the growing-order time-derivatives of the pristine data. By using different types of examples, we demonstrate how TiDe spaces are well- suited embeddings for information extraction, which hold considerable advantages in terms of complementarity of the information contained in the various TiDe's dimensions and that do not need dimensionality reduction approaches to extract physically-interpretable information.
This makes TiDe spaces ideal to study dynamical systems of phenomena of different types/scales and to unravel their internal complexity by analyzing their data.

In the specific test cases considered here, the most informative improvements arise by combining together the lowest-order derivatives. As shown in both MD benchmarks (Fig.~\ref{fig:fig2}) and experimental colloidal trajectories (Fig.~\ref{fig:fig5}), higher-order derivatives contain progressively less-and-less independent/complementary information. This indicates that the systems studied herein are quite low-dimensional in the TiDe spaces explored herein (at most 2-3-dimensional). 
However, when studying a complex system, its effective TiDe
dimensionality is unknown \textit{a priori} and may vary depending on
the system's complexity.
Since the TiDe framework can be formally extended up to arbitrarily high derivative orders, its application to other dynamical systems may reveal different scenarios. In particular, systems with sharper transitions or richer high-frequency features may benefit from the inclusion of higher derivatives (capturing patterns in e.g. accelerations, jerks, snaps, etc.). Thus, while our examples point to the predominance of low-order terms, the generality of TiDe leaves room for case-dependent optimization. The important requisite for TiDe analyses is just to start from an overestimation of the TiDe dimensions: as e.g. in the demonstrative ice/water coexistence case used herein, where the TiDe extends up to the 11$^{th}$ dimension (i.e., up to $D^{10}$). This guarantees to capture all information contained in, and practically extractable from, the time-derivatives, while the less useful ones will simply not contribute any additional information, as shown in Fig.~\ref{fig:fig4}.  

The robustness of our conclusions has been quantitatively assessed by running different types of complementary analyses. The Information Imbalance analyses of Fig.~\ref{fig:fig4} showed that the lowest order time-derivatives contain most of the information globally stored in the TiDe space (information present up to $\ddot D$). This was also confirmed via the time-lagged independent component analysis (tICA) shown in Figs.~\ref{fig:fig4}E,F. 
Comparison of the results obtained with the tICA analysis with those of Fig.~\ref{fig:fig2} and Fig.~\ref{fig:fig4}A-D also demonstrates that extracting information from such TiDe spaces does not require prior dimensionality reduction (which -- as it is the case of tICA in this specific case study -- may also lead to information loss).

The application of our approach to study the experimental Quincke roller trajectories of Fig.~\ref{fig:fig5} demonstrates how easy it is to use TiDes also in different scale systems, proving the versatility of the framework. The data of Fig.~\ref{fig:fig5} show how navigating TiDe spaces may provide useful physically-interpretable information for systems about which little is known a priori in a rather simple way, also for higher-scale systems that are monitored experimentally
While in our case studies lower-order derivatives proved most informative, this might not be always the case. Nonetheless, the TiDe framework presented herein is flexible enough to adapt to different systems, where higher derivatives might play a stronger role. Taken together, these findings establish TiDe as a robust and broadly applicable framework for extracting and interpreting the information contained in dynamically complex systems of different types and scales. Its combination of simplicity, adaptability, and interpretability makes TiDes a promising tool for the study of complex systems and phenomena in general, and for maximizing information extraction from their data.

\section*{Materials and Methods}
\subsection*{Systems trajectories}

\subsubsection*{Water/Ice coexistence}
The data reported in Figs.~\ref{fig:fig1} and \ref{fig:fig2} were obtained from a 50~ns MD trajectory of an atomistic system consisting of 2048 TIP4P/ICE water molecules~\cite{abascal2005potential}. The simulation was performed in NPT conditions, at a pressure of $P=1$~atm, starting from an initial configuration where half of the molecules were in the liquid phase and the other half arranged in a hexagonal ice lattice, and at the melting temperature ($T = 268$~K). Further details on the model and simulation setup are provided in~\cite{caruso2023timesoap}. Particle positions were saved along the 50~ns trajectory every $\delta t= 0.1$~ns. 

\subsubsection*{Quincke rollers}
The data shown in Fig.~\ref{fig:fig5} were obtained by analyzing the trajectories of 6921 Quincke rollers reconstructed from an experimental microscopy video~\cite{liu2021activity}. The recording spans a duration of 0.25~s over a microscopic field of $700 \times 700~\mu$m$^2$, containing polystyrene colloidal particles with an average diameter of $\sim 10~\mu$m. Particle positions in the $xy$ plane were tracked under a weak perpendicular DC electric field, which induces collective dynamics such as waves spanning the entire system. The $x$- and $y$-coordinates of all particles were extracted for 312 consecutive frames using in-house code~\cite{becchi2024layer} in combination with the Trackpy algorithm~\cite{allan2025trackpy}, yielding trajectories with a temporal resolution of $\delta t = 0.8$~ms. 

\subsection*{Descriptors calculation}
\subsubsection*{LENS}

LENS (Local Environments and Neighbors Shuffling) is a general-purpose, time-dependent descriptor designed to quantify local dynamical changes in many-body systems -- ranging from atoms and molecules to colloids, agents, or other interacting units. 

For each unit, LENS records the set of neighbors within a fixed cutoff radius at every time step, thereby generating a fingerprint of its local environment. The LENS value for a particle $i$, between two consecutive stored time points $t$ and $t+\delta t$, is defined as the normalized symmetric difference between the corresponding neighbor lists $C_i^t$ and $C_i^{t+\delta t}$: 

\begin{equation}
 \delta_i^t \equiv
 \frac{\bigl|C_i^t\cup C_i^{t+\delta t}\bigr|-\bigl|C_i^t\cap C_i^{t+\delta t}\bigr|}
 {\bigl|C_i^t\bigl| + \bigl|C_i^{t+\delta t}\bigr|}
\end{equation}

This reshuffling score ranges from $0$ (no change in neighbors) to $1$ (completely new neighbors), directly quantifying the degree of local reorganization. Time-series of $\delta$-values across all units provide a system-agnostic way to identify distinct dynamical domains, capture transitions (e.g., from solid-like to liquid-like states), and detect rare local events such as defect formation or particle diffusion. 
A detailed description of the method can be found in Ref.~\cite{crippa2023detecting}. LENS time-series were computed using the \texttt{dynsight} software~\cite{martino2025dynsight}. 

\subsubsection*{Local velocity alignment $\varphi$}

To characterize the collective dynamics of Quincke rollers, we used the local velocity alignment descriptor (Fig.~\ref{fig:fig5}A), $\varphi_i$, defined for each particle $i$ as  

\begin{equation}
    \varphi_i \equiv \frac{1}{n_c^i} \sum_{j=1}^{n_c^i} 
    \frac{\vec{v}_i \cdot \vec{v}_j}{|\vec{v}_i| \, |\vec{v}_j|}
\end{equation}
where $n_c^i$ is the number of neighbors of particle $i$ within a cutoff distance $r_c$, and $\vec{v}_i$ and $\vec{v}_j$ are the instantaneous velocities of particle $i$ and its neighbor $j$.  

This descriptor measures the degree of orientational alignment between the motion of particle $i$ and that of its neighbors: 
\begin{itemize}
    \item $\varphi_i = 1$ indicates perfect alignment, where $i$ and its neighbors move in the same direction, as would occur within the coherent core of a traveling wave. 
    \item $\varphi_i = -1$ corresponds to perfect anti-alignment, where $i$ moves in the exact opposite direction to its neighbors. 
    \item $\varphi_i \approx 0$ signifies orthogonal or uncorrelated motion, characteristic of regions where local dynamics are disordered, with no preferred directionality. 
\end{itemize}

Because it encodes how strongly particles align with their local environment, $\varphi$ is highly sensitive to the onset and propagation of collective excitations. For example, a sharp increase in $\varphi$ signals entry into the front of a propagating alignment wave, while a decrease toward zero marks the back of the wave where particles disalign.  

Thus, $\varphi$ provides a compact yet powerful descriptor: it translates microscopic particle velocities into a mesoscopic order parameter that directly reflects collective wave dynamics. When processed within the TiDe framework, $\varphi$ enables fine resolution of dynamical domains and transition regions, capturing not only stable states but also the transient fronts that drive system-wide reorganization. 
$\varphi$ time-series were computed using the \texttt{dynsight} software~\cite{martino2025dynsight}.

\subsection*{Analytical Methods}

\subsubsection*{Onion clustering}

Onion clustering is an unsupervised, iterative single-point time-series clustering algorithm designed to identify persistent environments within noisy time-series data~\cite{becchi2024layer}. The method operates through an iterative \emph{detect–classify–archive} loop, and requires a single parameter, a time resolution usually denoted with $\Delta t$. At each iteration, the most populated environment is detected in the distribution of signal values, represented by a Gaussian-like peak. All signal sequences of length $\Delta t$ that can be entirely assigned to this environment are then classified and removed from the dataset. By repeating this procedure, the algorithm progressively ``peels off'' layers of the data, until no further statistically relevant environments remain. 
At the end of the iterative process, a fraction of points can remain unclassified. These points typically correspond to fluctuations or transitions that are too short-lived to be statistically resolvable at the chosen temporal resolution. 

Importantly, Onion clustering is not restricted to a single analysis scale: the same procedure can be systematically repeated across multiple temporal resolutions $\Delta t$, ranging from the shortest sampling interval up to the full trajectory length. This provides, for each $\Delta t$, both the number of dynamically stable environments detected and the fraction of unclassified points, thereby revealing at which resolutions transient or metastable states become distinguishable.  

The method ensures interpretability, as each extracted environment is statistically validated and associated with a well-defined dynamical timescale. This feature makes Onion clustering broadly applicable to both simulated and experimental time-series, and particularly suited to scenarios where dynamical environments are partially overlapping, noisy, or hidden within complex fluctuations. 

\subsubsection*{Information Imbalance}

The Information Imbalance~\cite{glielmo2022ranking} is a statistical method to compare the relative informativeness of two data representations. Given a dataset with $N$ points, two different distance measures can be defined: $d_A$, computed from representation $A$, and $d_B$, computed from representation $B$. For each point $i$, its nearest neighbors are identified using $d_A$, and then the ranks of the same neighbors are checked according to $d_B$. If the neighborhoods defined by $A$ and $B$ largely coincide, then $A$ and $B$ encode similar information; if they disagree, one representation contains information not present in the other. 

The comparison is therefore not made between the raw values of $d_A$ and $d_B$, but between the neighborhood structures they induce. This makes Information Imbalance applicable to datasets of different dimensionality: whether a representation has one feature or many, the analysis always reduces to checking the consistency of neighborhoods across the two spaces. The imbalance $I_{\mathrm{IMB}}(A\to B)$ is an asymmetric measure bounded between 0 and 1, where $I_{\mathrm{IMB}}(A\to B)=0$ indicates equivalent information, $I_{\mathrm{IMB}}(A\to B)=1$ corresponds to statistical independence, and intermediate values capture partial or asymmetric containment of information. 

\subsubsection*{Time-lagged Independent Component Analysis}

Time-lagged Independent Component Analysis (tICA) is a dimensionality reduction method that identifies the slowest dynamical processes in time-dependent data. tICA finds the combination of input variables which maximize self time-correlation after a chosen lag-time $\tau$ \cite{molgedey1994separation, perez_identification, schwantes_improv, nuske_varia}. In other words, it extracts the directions along which the system evolves most slowly, filtering out fast fluctuations and noise. 

The procedure is based on constructing two covariance matrices from the data: one at lag-time $0$ and one at lag-time $\tau$. Solving a generalized eigenvalue problem between these matrices yields a set of new coordinates, called time-lagged independent components (tICs). 

In practice, the top few tICs provide a low-dimensional representation of the data that captures the essential long-timescale dynamics. This makes tICA especially useful for analyzing high-dimensional trajectories, where many microscopic fluctuations occur but the most relevant behavior is governed by slower collective motions. The method is linear, simple to compute, and has become widely used in the study of MD and other high-dimensional time-series~\cite{sittel_perspec}. tICA was performed using the \texttt{dynsight}~\cite{martino2025dynsight} and deeptime~\cite{hoffmann2021deeptime} python packages. 

\subsubsection*{Time-series filtering: pre–processing and low–pass filter}

All descriptor time-series were first prepared to mitigate high–frequency noise before computing time derivatives. Specifically, we applied a zero–phase Butterworth low–pass filter of order~4, implemented via forward–backward filtering (using the scipy Python package~\cite{virtanen2020scipy}). Zero–phase filtering avoids phase distortion by running the filter forward and backward in time, which effectively doubles the magnitude response order while canceling phase delay. 

To pick a physically sensible cutoff frequency, we computed the discrete Fourier transform of each signal and summed magnitudes across trajectories, then chose the smallest frequency at which the spectrum fell below a fixed fraction (here, $20\%$) of its maximum. This yields a data–driven cutoff that removes the noise–dominated high–frequency tail while preserving the dominant dynamical content~\cite{zhang_fourier}. The cutoff was always set strictly below the Nyquist frequency to prevent aliasing. 

To suppress filtering artifacts, we removed the first and last ten frames of each trajectory after low-pass filtering. Discrete differentiation is then performed, see Eq.~\ref{eq:eq1}. Since this process reduces the length by one frame at each derivative order, $D^{k}$ has $T-k$ frames if $D^0$ has $T$. To align all orders on a common grid, we trimmed the first $N-k$ frames of each $D^k$, yielding a final length of $T-N$, where $N$ is the maximum derivative order. Each time index $t$ in this aligned dataset is thus associated with a consistent vector $\left(D^0(t),D^{(1)}(t),\dots,D^{(N)}(t)\right)$, ensuring dimensional coherence in the TiDe construction.

This filtering and derivation workflow suppresses the well–known amplification of high–frequency noise caused by numerical differencing, stabilizes clustering, and preserves the physically meaningful low–frequency content of the dynamics. By computing derivatives on already denoised time-series, the analysis avoids spurious contributions that would otherwise arise if derivatives were applied directly to the raw, unfiltered time-series. 

\section*{Data Availability}
The codes and the full parameters setup used to perform the analysis conducted are available at \url{https://github.com/GMPavanLab/Information-Learning-in-TiDeS}.
Complete data and materials pertaining the trajectories employed herein will be available at a definitive Zenodo link upon acceptance of the final version of this paper. Additional information is available from the corresponding author upon reasonable request.

\section*{Author Contributions}
G.M.P. conceived and supervised this work. D.D. and M.B. developed and performed the analyses. All authors analyzed the data and contributed to the writing and the revision of the paper.

\section*{Acknowledgments}
G.M.P. acknowledges the support received by the European Research Council (ERC) under the Horizon 2020 research and innovation program (grant agreement no. 818776 - DYNAPOL). This work has also been supported by FAIR (Future Artificial Intelligence Research) and by the European Union Next-GenerationEU [Piano Nazionale di Ripresa e Resilienza (PNRR) – mission 4, component 2, investment 1.3 – D.D. 1555 October 11, 2022, PE00000013, CUP: E13C22001800001].

\section*{Conflict of Interest}
The authors have no conflicts to disclose.

% Bibliography
\bibliographystyle{unsrtnat}
\bibliography{bibliography}

%\end{document}

\clearpage
\onecolumn

\setcounter{figure}{0}
\renewcommand{\thefigure}{S\arabic{figure}}

\setcounter{table}{0}
\renewcommand{\thetable}{S\arabic{table}}

\setcounter{equation}{0}
\renewcommand{\theequation}{S\arabic{equation}}

%\usepackage{caption}
%\DeclareCaptionFont{figcapfont}{\fontsize{8pt}{10.8pt}\selectfont}
%\AtBeginDocument{\captionsetup[figure]{font=figcapfont,labelfont=bf}}

%\begin{document}

%\title{The Information Contained in Time-Derivatives Spaces}
%\title{Supporting Information for\\Physically-Relevant Information Learning in High-Dimensional Time-Derivatives Spaces}
% Building and navigating HDTDS 
% The information contained in HDTDS 
% Extracting information from HDTDS 
% Information search into HDTDS 

%\author[1]{Domiziano Doria}
%\author[1]{Matteo Becchi}
%\author[1]{Giovanni M. Pavan\thanks{Corresponding author: giovanni.pavan@polito.it}}
%\affil[1]{Department of Applied Science and Technology, Politecnico di Torino, Torino 10129, Italy}
%\date{\today}

%\maketitle
\newpage
\begin{center}
    {\LARGE\bfseries Supporting Information\par}
\end{center}

\vspace{1cm}
\subsection*{Supporting Figures}

\begin{figure}[ht!]
\centering
\includegraphics[width=\textwidth,keepaspectratio]{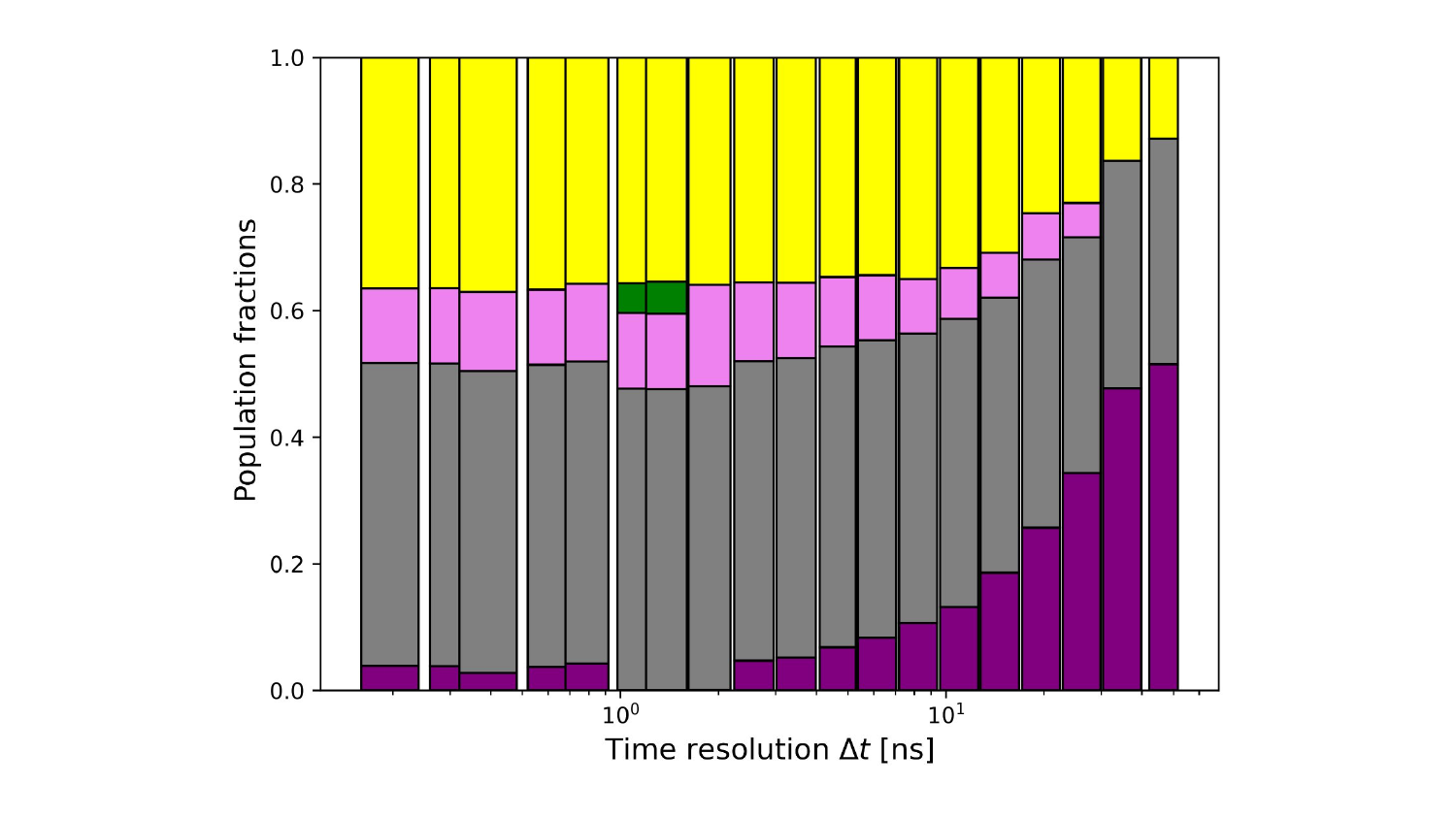}
\caption{Population fractions for the water--ice coexistence system as
a function of the Onion-clustering resolution $\Delta t$. The height of
each colored segment gives the fraction assigned to bulk ice (gray),
interface 1 (pink), interface 2 (green), or liquid water (yellow).}
\label{fig:s1}
\end{figure}
\newpage

\begin{figure}[ht!]
\centering
\includegraphics[width=17cm,keepaspectratio]{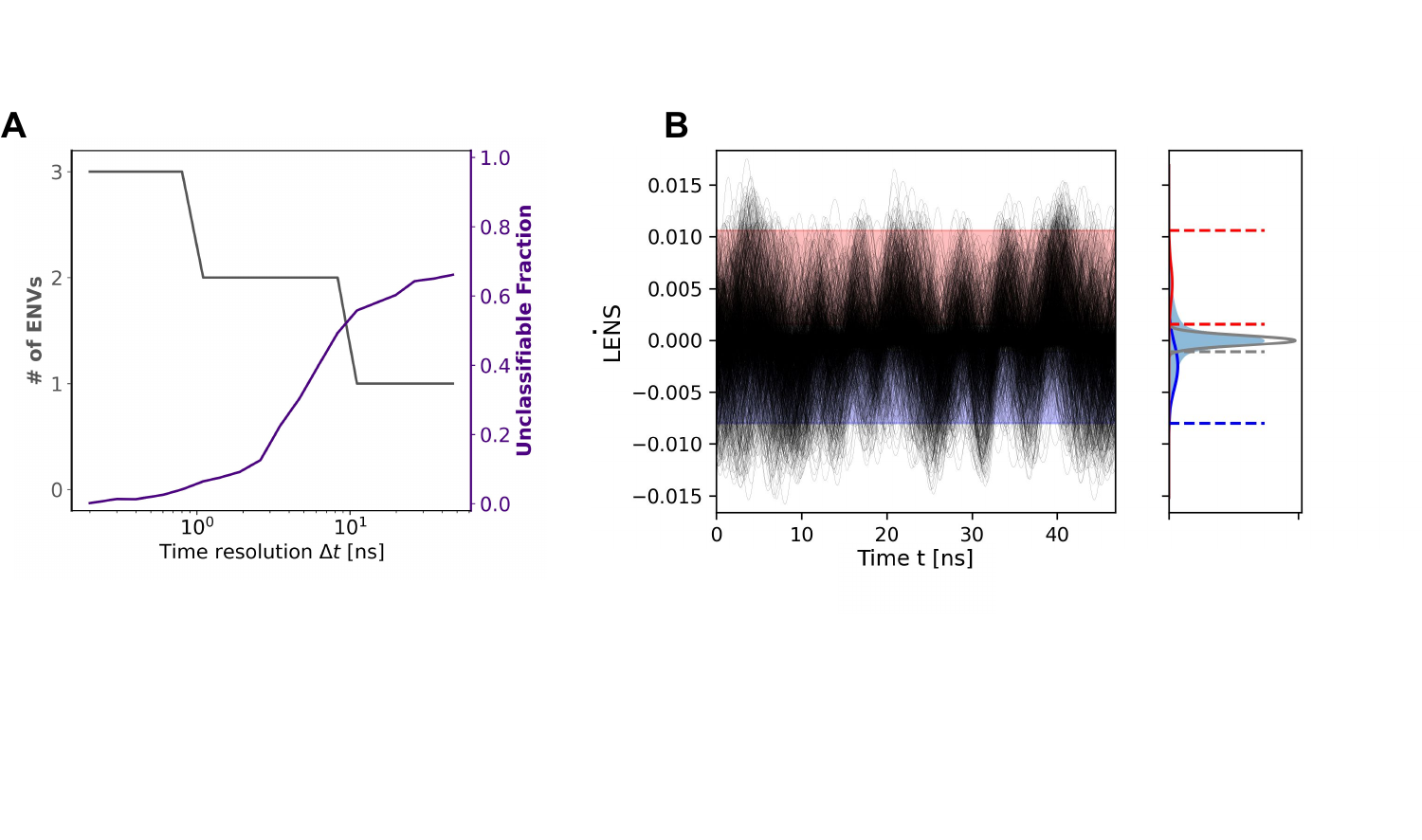}
\caption{Analysis of the $\dot{LENS}$ time-series alone.
(A) Onion-clustering resolution plot showing the number of detected
environments (gray, left axis) and the unclassifiable fraction
(violet, right axis) as functions of $\Delta t$.
(B) Clustered Time-series at $\Delta t= 0.4~ns$. Three separated clusters resolved: one centered in $\dot{LENS}=0$, containing the noises of all dynamically-distinct micro-clusters (gray) plus two symmetrical ones at positive (red) and negative (blue) $\dot{LENS}$ values.
}
\label{fig:s2}
\end{figure}
\newpage

\begin{figure}[ht!]
\centering
\includegraphics[width=\textwidth,keepaspectratio]{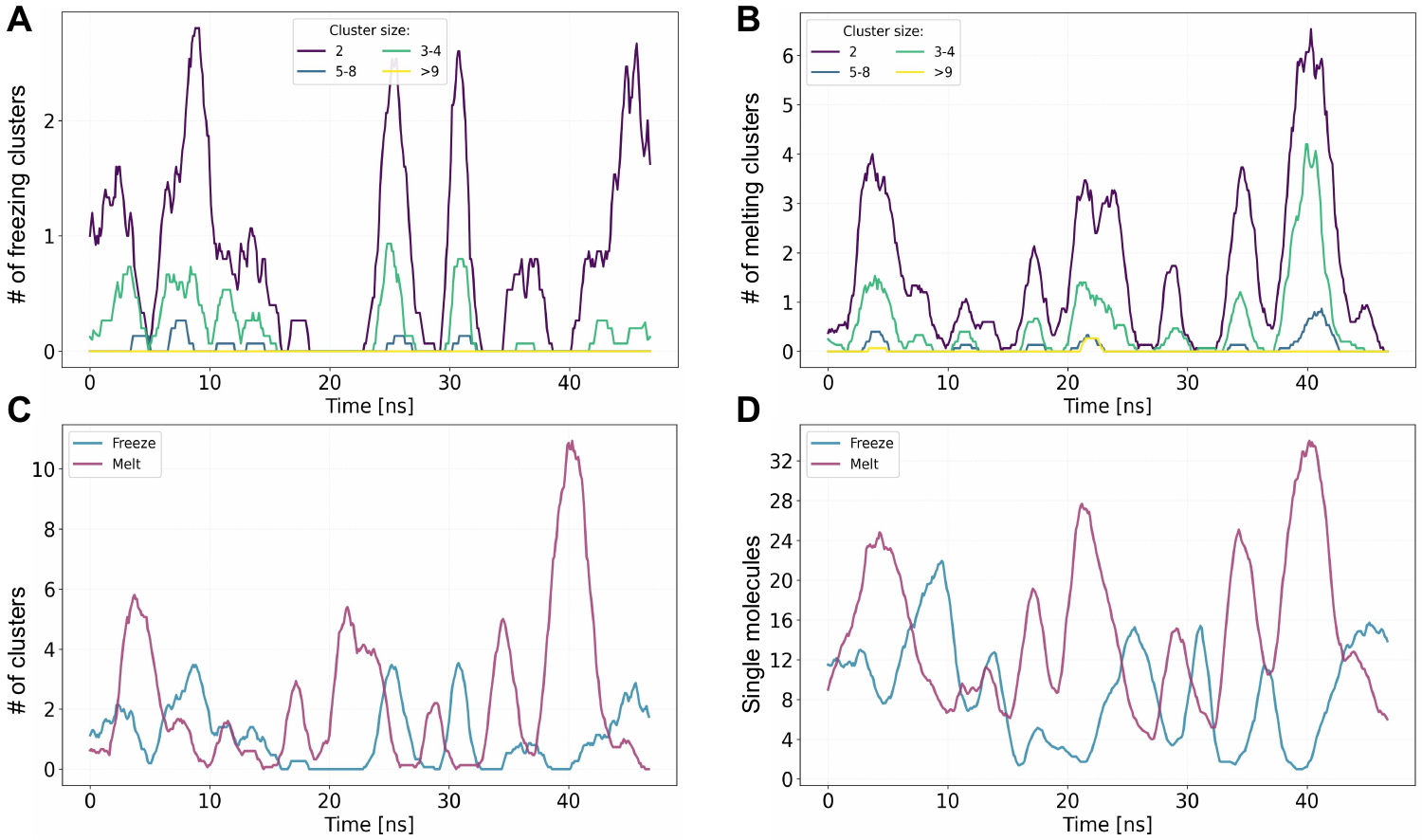}
\caption{
(A) Number of freezing clusters over time, grouped by cluster size and grouped into logarithmic cluster-size bins for clarity: size $=2$ (violet), size $3$--$4$ (green), size $5$--$8$ (blue), and size $>9$ (yellow).
(B) Number of melting clusters over time, grouped by cluster size and merged on a logarithmic scale for clarity: size $=2$ (violet), size $3$--$4$ (green), size $5$--$8$ (blue), and size $>9$ (yellow).
(C) Total number of clusters over time in the melting (pale red) and freezing (pale blue) environments.
(D) Total number of molecules outside clusters over time in the melting (pale red) and freezing (pale blue) environments.
}
\label{fig:s3}
\end{figure}
\newpage
\begin{figure}[ht!]
\centering
\includegraphics[width=\textwidth,keepaspectratio]{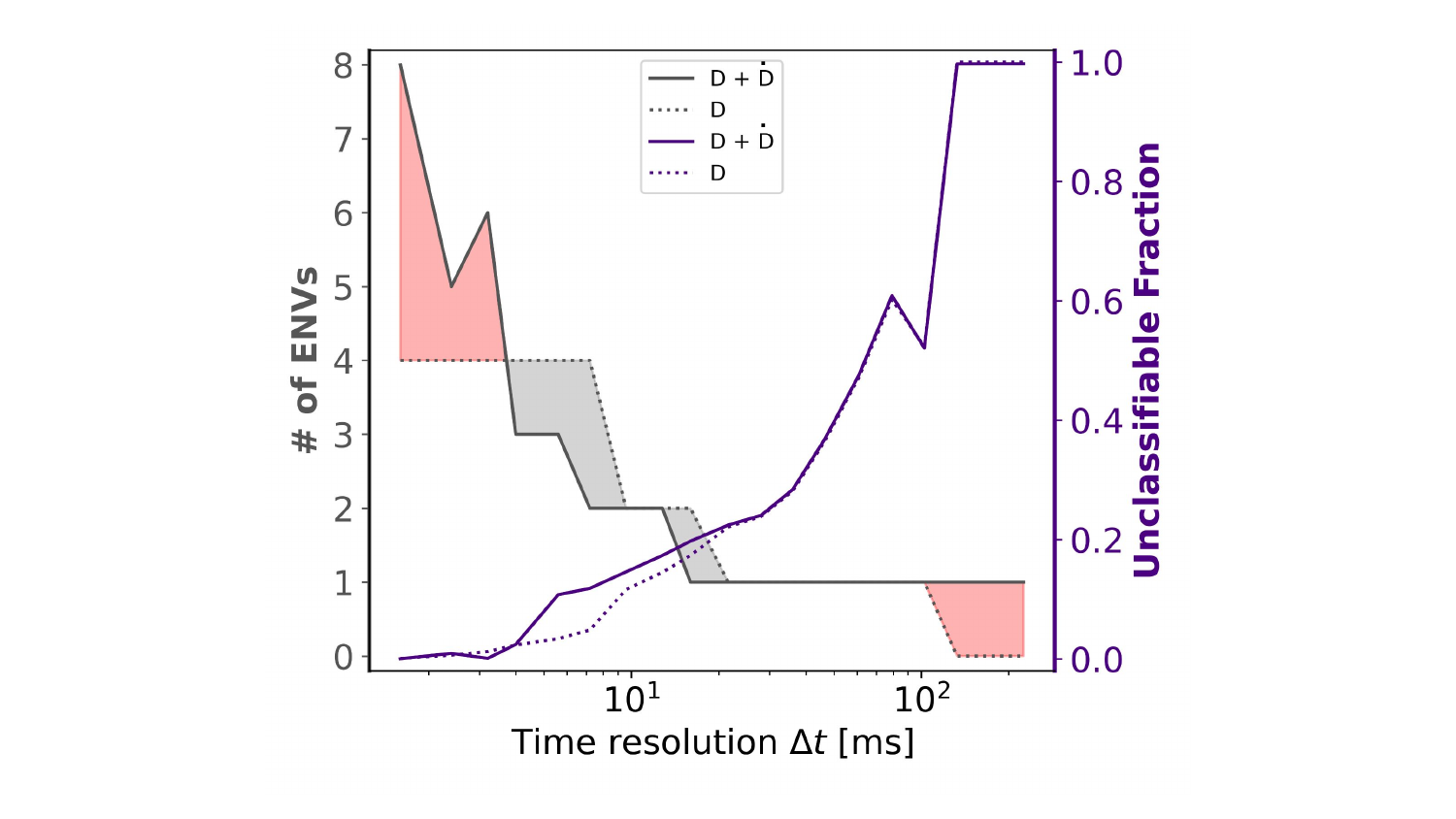}
\caption{Time-resolution onion plot referring to Quincke rollers' system. Comparison of Onion clustering using $\varphi$ alone (dotted lines)
and $(\varphi,\dot\varphi)$ (solid lines), showing the number of environments (gray) and the fraction of unclassified data points (violet) as a function of the Onion clustering time resolution $\Delta t$.
}
\label{fig:s4}
\end{figure}

\FloatBarrier

\subsection*{Supporting Movies}

\subsubsection*{Movie S1}
MD simulation of water/ice coexistence. Over time, the bottom region corresponds to ice (gray), the top region to liquid water (yellow), and two interfacial layers appear in between: the interface adjacent to the ice side (pink) and the interface adjacent to the liquid side (green).

\subsubsection*{Movie S2}
Time evolution in the two-coordinate $(LENS,\dot{LENS})$ space, clustered using Onion clustering. As in Movie~S1, clusters corresponding to bulk ice (gray), bulk liquid water (yellow), and the two interfacial layers (pink and green) are visible. In addition, freezing-transition molecules (blue) and melting-transition molecules (red) can be distinguished.

\subsubsection*{Movie S3}
Mapping of the clusters identified in the 2D space of Movie~S2 onto real space. This is closely related to Movie~S1, but in addition to the four main clusters, it also shows the spatial location of freezing-transition molecules (blue) and melting-transition molecules (red).

\subsubsection*{Movie S4}
Quincke roller particles moving from left to right, shown without any clustering applied to the trajectory.

\subsubsection*{Movie S5}
Clustered motion of Quincke rollers after applying the TiDe
representation and Onion clustering. The identified environments span
the maximum rate of increase of local alignment at the wave front (red),
intermediate alignment-increase environments (orange and yellow), the
wave core with high alignment and near-zero rate of change (green),
intermediate alignment-decrease environments after the wave passes
(cyan and azure), the maximum rate of decrease of alignment (blue), and
the random-orientation population (gray).

\end{document}